\documentclass{book}
\usepackage[dvips]{graphicx,graphics}
\usepackage{dcolumn}
\usepackage{bm}
\usepackage{epsfig}
\usepackage{amsmath}
\usepackage{fancyhdr}
\newcommand{\la}{\langle}
\newcommand{\ra}{\rangle}

\newcommand{\e}{\mathrm{e}}
\newcommand{\beq}{\begin{eqnarray}}
\newcommand{\eeq}{\end{eqnarray}}
\newcommand{\sbeq}{\begin{subeqnarray}}
\newcommand{\seeq}{\end{subeqnarray}}

\newcommand{\bl}{\biggl}
\newcommand{\br}{\biggr}
\newcommand{\mbfr}{\mathbf{r}}

\newcommand{\bfk}{\mbox{{\boldmath $k$}}}

\newcommand{\bfv}{\mbox{{\boldmath $v$}}}
\newcommand{\bfr}{\mbox{{\boldmath $r$}}}

\newcommand{\bfj}{\mbox{{\boldmath $j$}}}

\newcommand{\bfq}{\mbox{{\boldmath $q$}}}

\newcommand{\bfJ}{\mbox{{\boldmath $J$}}}

\newcommand{\bftheta}{\mbox{{\boldmath $\theta$}}}

\newcommand{\bfV}{\mbox{{\boldmath $V$}}}
\begin{document}

\title{Critical dynamics near  QCD critical point }
\author{Yuki Minami}
\maketitle

\maketitle

\chapter*{Abstract}

In this thesis, we study the critical dynamics near the QCD critical point.

Near the critical point, the relevant modes for the critical dynamics are 
identified as the hydrodynamic modes.
Thus, we first study the linear dynamics of them by the relativistic hydrodynamics.

We show that the thermal diffusion mode is the most relevant mode, whereas 
the sound mode is suppressed around the critical point.
We also find that the Landau equation, which is believed to be an acausal hydrodynamic equation,
has no problem to describe slowly varying fluctuations.
Moreover, we find that the Israel-Stewart equation, which is a causal one,
gives the same result as the Landau equation gives in the long-wavelength region.

Next, we study the nonlinear dynamics of the hydrodynamic modes
by the nonlinear Langevin equation and the dynamic renormalization group (RG).
In the vicinity of the critical point, 
the usual hydrodynamics breaks down by large fluctuations.
Thus, we must consider the nonlinear Langevin equation.
We construct the nonlinear Langevin equation based on 
the generalized Langevin theory. 
After the construction, we apply the dynamic RG to the Langevin equation
and derive the RG equation for the transport coefficients.

We find that the resulting RG equation turns out to be the same as that for the liquid-gas critical point
except for an insignificant constant. 
Consequently, the bulk viscosity and the thermal conductivity strongly diverge at the critical point.
Then, a system near the  critical point can not be described as a perfect fluid
by their strong divergences.

We also show that the thermal and viscous diffusion modes exhibit critical-slowing down 
with the dynamic critical exponents $z_{\rm thermal } \sim 3$ 
and $z_{\rm viscous} \sim 2$, respectively.
 In contrast, the sound mode shows 
critical-speeding up with the negative exponent $z_{\rm sound} \sim -0.8$.

\tableofcontents
\chapter{Introduction}

The quantum chromodynamics (QCD) is established as
the fundamental theory of the strong interaction.
Although the fundamental theory is established,
we can not study a strongly interacting matter based on the first principle
at finite density.
Thus, the QCD phase structure at finite density and temperature is not established \cite{Stephanov, qcdcp},
and determination of the structure is a fundamental problem.

\begin{figure}[!t]
    \begin{minipage}{1.0\hsize}
     \centering
     \includegraphics[width=\hsize]{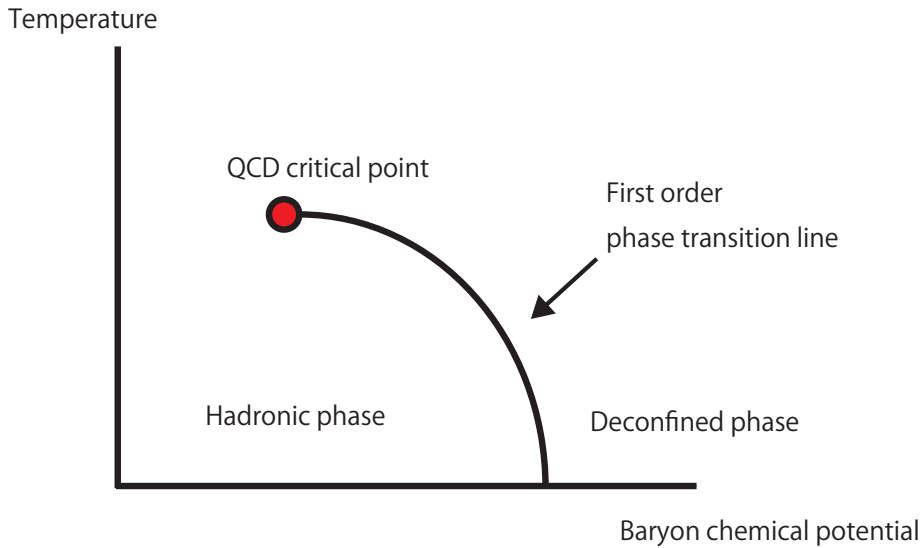}
       \caption{A schematic phase diagram of the QCD. }
     \label{fig: QCDpd}
    \end{minipage}
\end{figure}

Figure \ref{fig: QCDpd} shows a schematic phase diagram of a strongly interacting matter.
In low density and temperature region, we have the hadronic phase,
in which quarks and gluons are confined in hadrons.
On the other hand, in high density and temperature region,
the confinement breaks.
Then, deconfined quarks and gluons become relevant degrees of freedom.   
The phase boundary line, which separate the two phases, 
is predicted to be a first oder phase transition line 
by various effective models of the QCD \cite{qcdcp}.
This line exists in the finite density region, 
and then the first principle lattice QCD is not  available. 
In contrast,  transition along the temperature axis at zero density
is predicted to be crossover by the finite temperature lattice QCD \cite{Brown:1990ev}.
Namely, the transition is not associated with a thermodynamic singularity. 
The end point of the first order line is considered to 
be a second order transition point \cite{qcdcp, Fodor:2001pe}.
This point is called the QCD critical point \cite{Stephanov:2007fk}.

If the critical point exists, 
the correlation length of an order parameter, $\xi$, diverges
and thermodynamic quantities have singular behaviors at the point,
like the specific heat at the liquid-gas critical point.
By the singularity, the critical point is expected to be useful
for experimental probe of the QCD phase structure in the relativistic heavy ion collider \cite{bes, Stephanov:1998dy, ebef}.
Thus, the critical point attract the interest of many people.

It is known empirically that, by decreasing colliding energy, 
 the chemical potential of the created matter increases.  
 Then, by varying the colliding energy, we can experimentally scan a part of the phase diagram.
Such experiment is called the beam energy scan  program and now ongoing \cite{bes}.  
Several experimental signatures have been suggested based on the critical divergence\cite{Stephanov:1998dy, Kitazawa:2011wh}.
For example, the baryon number fluctuation is predicted to be enhanced, as
 $\la (\delta N) \ra \sim \xi^2$, near the critical point\cite{nongaussian}.
Thus, the baryon number fluctuation is expected to have the non-monotonic dependence
on the colliding energy, if the created matter passes near the critical point.

The static critical phenomena (namely, time-independent one)
have been strenuously studied.
Consequently, the order parameter, more appropriately, the critical mode\cite{reichl}, 
is now identified as a linear combination of the chiral condensate $\sigma$ and the baryon number density $n$
\cite{fujii}, if the critical point exists.  
Moreover, the static universality class is identified as the class of 3d Ising model, $Z(2)$\cite{Stephanov:2007fk}.
We note that the chiral symmetry is explicitly broken by the finite quark mass,
and thus $\sigma$ couples to $n$.

In contrast,  dynamic critical phenomena (for example,
critical-slowing down or divergence of transport coefficients) have not been fully studied.
The critical-slowing down is the phenomenon that the life time of the order parameter diverges at the critical point.
The dynamic critical phenomena are of a long-time scale.
 Thus, long-living modes (slow modes) are the relevant modes for the critical dynamics\cite{mazenko}.
 What are the slow modes near the QCD critical ?
 The slow modes are now considered as the hydrodynamic modes coming from the fluctuations of  conserved densities:
 the baryon number $n$ and the energy-momentum $T^{\mu \nu}$\cite{fujii, son}.
The slow dynamics of the order parameter, which is the linear combination of $\sigma$ and $n$,
is governed only by the baron number fluctuation.
Thus, the chiral condensate would be irrelevant.

Although the relevant modes are specified,
 dynamics of them has not been studied.
Specifically, the coupling between $\delta n$ and $\delta T^{\mu \nu}$ is not taken into account
in the earlier study\cite{son}.
Thus, in this thesis, we shall first study the linear dynamics of them
by the  relativistic hydrodynamics. 
Here, the important point is that the long-time behavior of the conserved densities, $n$ and $T^{\mu \nu}$,
is basically given by the hydrodynamics. 
As relativistic hydrodynamic equations,
we use the Landau equation and the Israel-Stewart equation.
The Landau equation\cite{landau} is believed to be an acausal hydrodynamic equation\cite{hiscock}.
However, we show that the equation has no problem to describe the hydrodynamic modes.
We shall also find that the Israel-Stewart equation\cite{is}, which is a causal equation, gives 
the same result as the Landau equation gives on the long-time and long-distance scale.   
Furthermore, we shall show that the actual slow modes is three: the thermal, 
 viscous, and  sound modes. 
We also find the relative importance of them;
the thermal mode is the most relevant, whereas the sound mode is suppressed 
around the critical point.

Furthermore, 
some authors suggested a divergence of the bulk viscosity at  the QCD critical point \cite{karsch}.
But, its validity is controversial
\cite{Romatschke:2009ng, moore, sasaki}; for example,
the limiting operation in the Kubo-formula may not be correct \cite{Romatschke:2009ng}, and 
a study by the relativistic Boltzmann equation \cite{sasaki} 
shows that the bulk viscosity is finite at the critical point.
Thus, it is still not obvious whether 
 the transport coefficients will diverge or not at the QCD critical point.

In fact, as is known in condensed matter physics, 
the critical divergence of the transport coefficients is a common phenomenon at a critical point,
  such as  at the liquid-gas critical point,
and originate from a universal mechanism; 
nonlinear interactions of {\em slow modes}  
cause the divergence \cite{kawasaki1, mori:fujisaka}.
This implies that fast modes, or microscopic processes as described by like
the Boltzmann equation, would not contribute 
to the critical divergence of these quantities, if any.
The dynamic renormalization group (RG) theory \cite{onuki, mazenko}
is a  standard technique for the critical dynamics,
which systematically incorporate the macroscopic nonlinear interaction 
causing the  divergence of transport coefficients.
In this theory, 
We must construct a nonlinear Langevin equation 
to describe the nonlinear interaction of the slow modes. 
The construction goes as follows.
First,  We identify the slow variables, which label a state on the long-time and long-distance scale.
Next, the thermodynamic potential for the slow variables is constructed
to determine the static property of the system.  
Finally, the streaming terms, which cause the dynamic-nonlinear interactions,
and the kinetic coefficients are determined
for respectively describing time reversible and irreversible changes of the slow variables.
We note that the streaming term is absent in the simple  Brownian motion.

The general theory of the critical dynamics as described above
tells us that an essential ingredient is 
to  properly construct the  nonlinear Langevin equation
for the critical dynamics.     
As far as we know, this is the first attempt for the QCD critical point.
Our construction of the Langevin equation  
is based on the generalized Langevin theory, so-called the Mori theory \cite{mori, mori:fujisaka}, 
and the {\em relativistic} hydrodynamics, 
because the slow modes are identified as the hydrodynamics modes 
\cite{fujii,son,Minami:2009hn};
we construct the streaming terms from continuity equations and 
the potential condition, which is a general condition for streaming terms \cite{onuki, mazenko}.
Also, we use the thermodynamic potential for 
the 3d Ising system as that for the QCD critical point
because the static universality class is the same as 3d Ising class \cite{fujii,son,onukin}.
Finally, we determine the kinetic coefficients 
from a relativistic hydrodynamic equation, here  
the Landau equation \cite{landau} used.
Consequently, we shall show that 
the Langevin equation differs from it for the liquid-gas critical point by relativistic effects,
although the dynamic universality class of the QCD critical point is conjectured as of the liquid-gas critical point \cite{son, moore}.  

After such construction, 
we apply the dynamic RG to the Langevin equation and derive the RG equations for the transport coefficients.
Consequently, to our surprise,
these RG equations turn out to be the same as for the liquid-gas critical point 
except for a irrelevant constant,
although the Langevin equations are different.    
Therefore, the bulk viscosity and the thermal conductivity strongly diverge 
and can be more important than the shear viscosity near the QCD critical point.
We shall also show that the thermal and viscous diffusion modes exhibits critical slowing down,
whereas the sound mode critical speeding up.  

This thesis is organized as follows. 

In Chap.\ref{sec:tocd}, we review the general theory of critical dynamics.
This thesis is based on the general theory.
Specifically, we first give the projection operator method.
By this method, 
we can systematically decompose  any dynamic variables into a slowly varying motion and a rapid one.
Namely, we  can extract the relevant motion for the critical dynamics.

We also give the statistical basis of  coarse-grained equations of motion, namely,  
the Langevin equation and the hydrodynamic equation.
We note that a microscopic theory  is difficult to describe the long-time effect
although the dynamic critical phenomena are of the long-time scale.
Thus, the coarse-grained equation is  relevant.

After the projection operator method, 
we briefly explain  the typical dynamic critical phenomena:
the critical-slowing down and the critical divergence of  transport coefficients.
We also give the concept of the dynamic RG, and the critical and hydrodynamic regimes.
Near the critical point, the macroscopic scale is divided  into the two regimes.

Moreover, we also give the earlier studies on the slow modes near the QCD critical point.  
 
In Chap.\ref{sec:sdaqcp}, we study the linear dynamics 
of the slow modes by the relativistic hydrodynamics.

In Chap.\ref{sec:drgqcd},  we study the nonlinear dynamics 
of them by the dynamic RG.

In the final chapter, we give summary and concluding remarks.

\chapter{Theory of critical dynamics}
\label{sec:tocd}

Here, 
we provide the general theory of critical dynamics, which is developed in condensed matter physics.

We also give earlier studies \cite{fujii, son} on the QCD critical point,
which shows the relevant variables are the hydrodynamic ones.   

\section{Projection operator method}

The Mori's projection operator method gives the microscopic basis of a Langevin equation
\cite{mori, mori:fujisaka}.
By this method, we can formally extract dynamics in the long-time scale,
which is relevant for critical phenomena.   
The Langevin equation has been widely 
used to study critical dynamics.

\subsection{Linear Langevin equations}
\label{sec:lle}

Here,  we derive a linear Langevin equation from a microscopic equation 
by the Mori theory\cite{mori, onuki}.

Now,  let us consider a classical many body system, for simplicity. 
A generalization to a quantum system is straightforward. 
For the classical system, a time evolution of an arbitrary dynamic variable
is given by the Liouville equation
\footnote{If we replace the Liouville equation with the Hisenberg equation,
the following arguments  are valid for a quantum system. }
:
\begin{equation}
\frac{\partial}{\partial t} B(t) = \{ B(t), H \}_{\rm PB}. \label{eq: liouville},
\end{equation}
where $B(t)$ is a dynamic variable in time $t$, 
$H$ a microscopic Hamiltonian and $\{ , \}_{\rm PB}$ the Poison bracket.
Introducing the Liouville operator, $i{\cal L}$, as
\begin{equation}
i{\cal L} B(t) \equiv \{ B(t), H \}_{\rm PB}.
\end{equation}
we can formally solve Eq.(\ref{eq: liouville}) as 
\begin{equation}
B(t) =e^{i {\cal L} t}B(t=0). \label{eq: te}
\end{equation}
In the following, we shall decompose the time evolution Eq.(\ref{eq: te}). 

First, let us represent a set of slowly varying variables (slow variables) as $\{A_j(t)\}$.
The slow variables label a macroscopic state and describe a time evolution on a macroscopic scale.

Next, we define the linear projection operator ${\cal P}$ as 
\begin{equation}
{\cal P}B(t) =\sum\limits_{j k} \la B(t) A_j \ra \chi_{j k}^{-1} A_k. 
\end{equation}
Here, $\la ... \ra$ is the equilibrium-statistical average,  
$A_j$ an initial value of the slow variable, namely $A_j \equiv A_j (t=0)$, 
and $\chi_{j k}^{-1}$ the inverse of the correlation $\chi_{j k} \equiv \la A_j A_k \ra $. 
In the following, we denote the initial values of the slow variables  
without the argument $t$.

The operator ${\cal P}$ extracts a slowly varying motion from arbitrary dynamical variables. 
Also, we define the orthogonal operator, as ${\cal Q} \equiv 1 - {\cal P}$.

Now, we use the operator identity, which is valid for arbitrary $i{\cal L}$ and ${\cal P}$\cite{onuki},
 \begin{equation}
\frac{\partial }{\partial t} e^{i{\cal L}t}=e^{i {\cal L} t}{\cal P}i {\cal L}
+\int^t_0dt^{'}e^{iL(t-t^{'})}{\cal P}i {\cal L} e^{{\cal Q}i{\cal L}t^{'}}{\cal Q}i{\cal L}
+e^{{\cal Q}i{\cal L}t}{\cal Q}i {\cal L}. \label{eq: oi}
\end{equation}   
Multiplying Eq.(\ref{eq: oi}) by the initial values of the slow variables $A_j $,
we obtain the linear Langevin equation for $A_j (t) = \exp [i {\cal L} t] A_j $:
\begin{equation}
\frac{\partial}{\partial t} A_{j}(t) =
\sum\limits_k i\Omega_{j k}A_{k} (t)-\sum\limits_{k}\int^{t}_{0}dt^{'}\Gamma_{j k}(t^{'})A_k(t-t^{'})
+ f_j (t), \label{eq: gllangevin}
\end{equation}
without any approximations. 
Here, we introduced
\begin{eqnarray}
i\Omega_{j k} &&=\sum\limits_k \la \dot{ A_j} A_l\ra \chi_{l k}^{-1}, \label{eq: fmatrix}\\ 
f_j (t) &&=\exp{({\cal Q}i {\cal L} t)}{\cal Q}\dot{A_j}, \label{eq: lf} \\
\Gamma_{j k} (t) &&=\sum\limits_{l}\la f_j(t) f_l(0) \ra \chi_{l k}^{-1}, \label{eq: memory}
\end{eqnarray}
with $\dot{A_j}\equiv i{\cal L} A_j$.
The equation (\ref{eq: gllangevin}) has the following properties.
\begin{enumerate}
\item Eq.(\ref{eq: gllangevin}) is {\em the exact relation}. 
         Here, we only used the operator identity. 
\item The first term in the right-hand side is a time-reversible change.
\item The second term is a time-irreversible change.
         Also, this term depends on a past time value, $A_k(t-t^{'})$.
         $\Gamma_{j k} (k)$ is called a memory function.
\item The last term is a rapid motion and usually treated as a random noise.
\end{enumerate}

Now, we give a transport coefficient in this scheme.
If the time-scale of  the slow variables $A_j (t)$ and that of the noise $f_j(t)$ are well separated, 
we can assume that $A_k(t-t^{'})$ does not change among the correlation time of the memory function, 
Eq.(\ref{eq: memory}).
Namely, we approximate the time-irreversible term as 
\begin{eqnarray}
\int^{t}_{0}dt^{'}\Gamma_{j k}(t^{'})A_k(t-t^{'}) &&
       \sim \bl( \int^{\infty}_{0}dt^{'}\Gamma_{j k}(t^{'}) \br) A_k(t).                                                   
\end{eqnarray}
This approximation is called the Markov approximation.
We now introduce the linear transport coefficient as 
\begin{eqnarray}
L_{j k} =&&\sum\limits_{l} \bl( \int^{\infty}_0  dt \Gamma _{j l}(t) \br) \chi_{l k} , \\
         =&&\int^{\infty}_0 \la f_j(t) f_k(0) \ra . \label{eq:  trans} 
\end{eqnarray}
We see that the transport coefficient is given as the time correlation of the noises.
Finally, we obtain 
\begin{equation}
\frac{\partial}{\partial t} A_{j}(t) =
\sum\limits_k i\Omega_{j k}A_{k} (t)-\sum\limits_{k i}L_{j i}\chi^{-1}_{i k}A_k(t)
+f_j(t). \label{eq: llangevin}
\end{equation}
We see that Eq.(\ref{eq: llangevin}) loses the memory effect.

An important point is that the noise, $f_j (t)$, implicitly includes nonlinear terms of the slow variables.
Namely, $f_j(t)$ is not orthogonal to the nonlinear terms:
\begin{eqnarray}
\la f_j(t) A_k \ra &&=0,  \\
\la f_j(t) A_k A_l\ra && \neq 0 .
\end{eqnarray}
This originates from that  ${\cal P}$ is {\em a linear projection operator}.
Therefore, if we can not neglect the nonlinearity, $f_j(t)$ can not be treated as a noise.

Moreover, from Eq.(\ref{eq: trans}), 
we see that the nonlinear terms contribute to the linear transport coefficients.
This contribution causes the critical divergence of the transport coefficients near a critical point.   
Near the critical point, fluctuations become large.
Therefore, we can not neglect the nonlinear  fluctuations 
and must consider a nonlinear Langevin equations as a basic equation
for critical dynamics.

\subsection{Nonlinear Langevin equations}
\label{sec:nle}

Here, we derive the nonlinear Langevin equation 
by the nonlinear projection operator \cite{mori:fujisaka, onuki}.  

To define the nonlinear projection operator, 
we first introduce the following delta functional as
\begin{equation}
g(A,a)\equiv \prod\limits_j \delta (A_j - a_j ),
\end{equation}
where $a_j$ are some initial values.
The equilibrium-statistical average of this gives the equilibrium-distribution function:
\begin{equation}
P_{\rm e q}(a) =\la g(A,a) \ra. 
\end{equation}

With this delta functional, we can define the nonlinear projection operator 
acting on any dynamic variables $B$ as 
\begin{equation}
{\cal P}_{\rm n l} B \equiv \la B g(A, a) \ra / P_{\rm eq}(a) \label{eq: nlp} .
\end{equation}
The physical meaning is simple.
We fix the slow variables $A_j$ at some values $a_j$ and 
average out the other degree of freedom.
In other words, we eliminate the fast variables and 
extract a slowly varying part that is determined by only the slow variables.

The important point is that the {\em nonlinear} projection on $A_j$
is identical to the {\em linear} projection on $g(A,a)$.
The linear projection on $g(A,a)$, which is ${\cal P}_g $, is given by
\begin{eqnarray}
{\cal P}_{ g} B &&= \int da da^{'} \la B g(A, a) \ra \la g(A,a)g(A,a^{'}) \ra^{-1} g(A, a^{'}), \\
                   &&=\la B g(A, a )\ra / P_{\rm eq}(a),\\
                   &&= {\cal P}_{\rm n l} B .
\end{eqnarray}
Here, we used the relations
\begin{eqnarray}
\la g(A,a)g(A,a^{'}) \ra &&= \delta (a-a^{'}) P_{\rm eq} (a), \\
\la g(A,a)g(A,a^{'}) \ra^{-1} &&= \delta (a-a^{'}) / P_{\rm eq}(a).
\end{eqnarray}
Namely, the linear projection ${\cal P}_g$ is equivalent to the nonlinear projection ${\cal P}_{\rm nl}$.
Hence, we can derive the nonlinear Langevin equation about $A_j$
from the linear Langevin equation about $g(A,a)$.
However, we leave the derivation to Appendix \ref{sec:a1}.

The resulting nonlinear Langevin equation with the Markov approximation
is  \cite{mori:fujisaka, onuki} 
\begin{equation}
\frac{\partial}{\partial t}A_j(t)=v_j(A)
-\sum\limits_{k} L_{j k}(A) \frac{\delta (\beta H(A) )}{\delta A_k}+\theta_j (t),
\label{eq: nllangevin}
\end{equation}
with $\beta$ being the inverse temperature.
Here, we  introduced 
\begin{eqnarray}
v_j(a)        &&=\la \dot{A}_j; a \ra , \label{eq: vja}\\
\theta_j(t) &&=\exp{[{\cal Q}_g i {\cal L}t ]} {\cal Q}_g \dot{A}_j(0), \\
L_{i k}(a)    &&= \int^{\infty}_0 dt\la \theta_i (t) \theta_k (0);a\ra ,
\end{eqnarray}
where $\la ... ; a\ra \equiv \la ...g(A,a)\ra/P_{\rm eq}(a) $ 
is the conditional average  in which $A_j$ is fixed at $a_j$.
Also, we defined the thermodynamic potential (or the effective potential) $H(A)$ as 
\begin{equation}
P_{\rm eq}(A) = \frac{1}{Z}\exp [-\beta H(A)], 
\end{equation}
where $Z$ is a normalization constant.

Now, we give physical meanings of Eq.(\ref{eq: nllangevin}).
\begin{enumerate}
\item The first and second terms are the slow motions and  nonlinear in $A_j$.
\item The first term, which is called the streaming term, gives a time-reversible change.
\item The second term gives a time-irreversible change. 
         $L_{i k}(a)$ is called the bare transport coefficient.
\item The last term is a fast motion 
         and treated as a stochastic variable obeying the fluctuation-dissipation relation
         \begin{equation}
         \la \theta_j (t) \theta_k(t^{'}) ;a \ra = 2 L_{j k}(a)\delta(t-t^{'}). \label{eq: fdr}
          \end{equation}
          In contrast to the linear case, $\theta_j(t) $ does not include the slow variables.
          Namely, the nonlinear terms of $A_j$ are explicitly extracted in the first and second terms. 
\end{enumerate}
Even for the QCD critical point, we may use the generalized Langevin equation, because 
 only the time-scale separation is assumed in the Mori theory.
Furthermore, we note that, by the time-scale separation, transport coefficients arises .

\subsection{On slow variables}
\label{sec:osv}

In the Mori theory, a choice of the slow variables plays a crucial role.
How we choose the slow variables?
On a macroscopic scale, the slow variables are given as conserved densities,
Nambu-Goldstone modes (NG modes), and order parameters \cite{mazenko}.
We now explain why they are slow.

First, let us consider the conserved density.
The important point is that
any conserved densities generally obey a continuity equation:
\begin{equation}
\frac{\partial n(\bfr , t)}{\partial t} = - \nabla \cdot \bfj (\bfr ,t) ,
\end{equation}
where $n$ is a conserved charge density and $\bfj$ is its current density.
Performing Fourier transformation about $\bfr$, we have
\begin{equation}
\frac{\partial n(\bfk , t)}{\partial t} = - i \bfk \cdot \bfj (\bfk ,t), \label{eq: continuity}
\end{equation}
where $\bfk$ is the wavenumber.
We see that
the time-change rate is proportional to the wavenumber.  
Thus, the conserved density is slow in the low-wavenumber region, namely, the macroscopic scale.

Next, we consider the NG modes.
The key is that  the NG modes generally have  gapless-dispersion relations. 
Namely, their dispersion relations are proportional  to the wavenumber or the square of that:
\begin{equation}
\omega (k) \propto k \mbox{ or } k^2,
\end{equation} 
where $\omega$ is a frequency.
Again, in the low-wavenumber region, we have the slow motion for the NG mode.

Finally, let us consider the order parameters.
In general, near a critical point, 
the relaxation of the order parameter is anomalously slow.
Such the dynamic critical phenomenon is called  the critical-slowing down.
Thus, the order parameter is slow near the critical point. 
We shall explain about the critical-slowing down in the next section.

Now, the important point on the slow variables is that  the low-wavenumber components are slow;
but the high-wavenumber components are fast.
Thus, we must restrict the wavenumber by the ultraviolet cutoff, $\Lambda$.
Then, the Langevin equation has the ultraviolet cutoff.

\section{Critical slowing down}

Here, let us illustrate the critical-slowing down by  a linear Langevin equation.
For example, we consider the case that a single-order parameter, $\phi$, is only the slow variable.
In this case, we have the following linear Langevin equation:
\begin{equation}
\frac{\partial \phi (t)}{\partial t} = \frac{L_{\phi \phi}}{\chi_{\phi \phi}}\phi (t) + f(t).
\end{equation}    
Here,  a reversible term is absent.
The reason is the following.
From Eq.(\ref{eq: fmatrix}), we have the relation \cite{mazenko},
\begin{equation}
i\Omega_{i j} =\la \{A_i,A_j \}_{\rm PB} \ra \mbox{ or } \la [ A_i, A_j ]/(i \hbar ) \ra . \label{eq: fmatrix2}
\end{equation}

Here, $ [ ..., ... ]/(i \hbar ) $ denotes commutation relation for a quantum system.
Therefore, if we have a single slow variable, the reversible term is generally absent.

Now, the important point is that  
 the susceptibility of the order parameter, $\chi_{\phi \phi}$, generally diverges at the critical point.
Thus, the relaxation of the order parameter exhibits  slowing down near the critical point\footnote{
As we shall see in the section\ref{sec:cdotc}, the transport coefficient, $L_{\phi \phi}$, also diverges.
However, the divergence of the transport coefficient is typically weaker than that of the susceptibility. }.

\section{Hydrodynamic and critical regimes}

Here, we give a valid region of the linear and nonlinear Langevin equations
in terms of the wavenumber and the cutoff\cite{onuki, HALPERIN:1969zza}.

Now, let us consider a non-equilibrium state near an equilibrium state.
In other words, the state fluctuating from the equilibrium state is considered.
We note that the critical point is defined on {\em the equilibrium phase diagram}.  
For such state,  the dynamic variables turn out to be fluctuations from the equilibrium state.
Then, if we consider a state far from the critical point, the fluctuations are small
and the linear Langevin equation suffices.

In terms of the wavenumber, the valid region is 
\begin{equation}
0 \leq k \ll a^{-1},
\end{equation}  
where $a$ is a microscopic characteristic length scale.
Thus, the ultraviolet cutoff of the linear Langevin equation,
$\Lambda_{\rm L}$, is chosen as
\begin{equation}
\Lambda_{\rm L} \ll a^{-1}, 
\end{equation} 
in the normal region.

In contrast to the normal region, near the critical point, the fluctuations become large.
Thus, we can not neglect the nonlinear fluctuations and 
must basically use the nonlinear Langevin equation.
However, if we restrict our interest to a much larger scale than 
the correlation length of the order parameter, $\xi$,
we can again use the linear Langevin equation even in the critical region.
On such scale, the information of the critical point is obscure and included in parameters, 
like the transport coefficients.
Namely, the cutoff for the linear theory must be chosen as 
\begin{equation}
\Lambda_{\rm L} \ll \xi^{-1},
\end{equation}   
in the critical region.

In contrast,  
the nonlinear Langevin equation can describe the nonlinear fluctuations
and thus  is valid even on a smaller scale than $\xi$.  
Then , we can choose the cutoff for the nonlinear Langevin equation as
\begin{equation}
\Lambda_{\rm NL} \ll a^{-1}, 
\end{equation}
even in the critical region.

The important point is that, near the critical point, 
the wavenumber regime is divided into the two regimes: hydrodynamic and critical regimes.

The hydrodynamic regime is 
\begin{equation}
0 \leq k  \ll \xi^{-1} .
\end{equation}   
In this regime, the linear theory is still valid
\footnote{The linear Langevin equation is sometimes called  hydrodynamic theory.
So,  this regime is called the hydrodynamic regime. }.
On the other hand, the critical regime is  
\begin{equation}
 \xi^{-1} \ll k \ll a^{-1}. 
\end{equation}      
In this regime, only the nonlinear theory is valid 
and fully reflects the information on the critical point.
 
We shall discuss the two regimes in terms of the dynamic RG in Sec.\ref{sec:DRG}.

\section{Critical divergences of transport coefficients}
\label{sec:cdotc}

The critical divergence of transport coefficients (or diffusion constants) is
a common phenomenon, for instance, 
at  the liquid-gas critical point, ferromagnetic transitions  and so on
\cite{onuki, mazenko}.
The important point is
that the critical divergence 
 originates from a universal mechanism; 
 macroscopic nonlinear fluctuations, namely, the nonlinear terms of slow variables,  
cause the divergence, which is implied in Sec.\ref{sec:lle} \cite{kawasaki1, mori:fujisaka}.

Now, 
we illustrate how the macroscopic nonlinear fluctuations
cause the critical divergence.
For an example, let us consider 
the thermal conductivity near the liquid-gas critical point \cite{kawasaki2}.
The thermal conductivity is given by the Kubo formula as follows,
\begin{equation}
\lambda = T^{-2}\int d{\mbfr}\int^{\infty}_0 dt \la q(\mbfr,t) q(0,0) \ra,
\label{eq:heat-kubo}
\end{equation}
where $q(\mbfr,t)$ and $T$ are the heat current and temperature, respectively.
The heat current $q(\mbfr,t)$  
consists of two parts:
one is due to a microscopic process as described by a microscopic theory, like
the Boltzmann equation, and the other is  
by the nonlinear fluctuations of macroscopic variables \cite{mori};
\begin{equation}
q = q_{\rm micro}+q_{\rm macro},
\end{equation}
where $ q_{\rm micro}$ and $q_{\rm macro}$ respectively denote
 the microscopic and macroscopic currents.
The macroscopic process causing the heat current is identified as
the entropy density convected by
fluid velocity fluctuation. Thus,
we have
\begin{equation}
q_{\rm macro}\sim \delta s \delta v,
 \label{eq: hc}
\end{equation}
where $\delta s$ and $\delta v$
 respectively denote the fluctuations of  the entropy density and  the fluid velocity.
The macroscopic current Eq. (\ref{eq: hc}) is of the second order in fluctuations 
and hence negligible far from the critical point.
However, it becomes the main part near the critical point,
because the fluctuations are enhanced there.
We see that Eq. (\ref{eq:heat-kubo}) 
now has the following form
\begin{equation}
\lambda \sim \lambda_{\rm micro}+\int d\mbfr\int_0^{\infty} dt 
\la\delta s(\mbfr,t) \delta v(\mbfr,t)\delta s(0,0)\delta v(0,0) \ra,
\label{eq: tc}
\end{equation}
where $\lambda_{\rm micro}$ is the thermal conductivity coming from $q_{\rm micro}$.
Recalling that the entropy density fluctuation is the order parameter for the liquid-gas critical point,
we see that the second term of Eq. (\ref{eq: tc}) diverges at the critical point.
This is the mechanism causing the critical divergence of transport coefficients.

Let us call the transport coefficients, such as $\lambda_{\rm micro}$, coming from  microscopic processes
the bare transport coefficients,
and those including the contributions from the nonlinear macroscopic fluctuations 
the renormalized ones.

Then, we need not to study the critical divergence of  transport coefficients by a microscopic theory 
because the divergence originates from only the macroscopic processes.
The dynamic RG \cite{onuki, mazenko, siggia, kg, onukin} is the 
standard theory treating such nonlinear macroscopic fluctuations.
In this theory, we must construct a nonlinear Langevin equation 
as a basic equation for the critical dynamics. 

Furthermore, we note that the earlier studies \cite{karsch, sasaki}, on 
the transport coefficients near the QCD critical point, treat the bare part.
Thus, those do not take into account the contribution from nonlinear macroscopic fluctuations.
In Chap.\ref{sec:drgqcd}, we shall study the renormalized part by the dynamic RG.

\section{Dynamic RG}
\label{sec:DRG}

Here, we give a conceptual aspect of the dynamic RG.
A technical aspect is well given in the textbook \cite{mazenko}.  

The general dynamic RG transformation usually 
consists of  two procedures, 
i.e., coarse graining and rescaling as in the
static RG transformation \cite{hhs, mazenko}.
However, as is shown in \cite{kg, onukin, onuki},
we can omit the rescaling, 
if we are interested in only  the critical exponents of transport coefficients.

The nonlinear Langevin has a ultraviolet cutoff $\Lambda_0$, 
which should satisfy the following inequality
\begin{equation} 
\xi^{-1} \ll \Lambda_0 \ll a^{-1}.
\end{equation}
Namely, at the starting point of the dynamic RG, we are in the critical regime.

Then, the  Langevin equation is coarse grained by
averaging over the high-wavenumber components
 of the slow variables $A_j(t)$ in the infinitesimal wavenumber shell,
\begin{equation} 
\Lambda-\delta \Lambda < k <\Lambda,
\end{equation}
for Eq. (\ref{eq: nllangevin}).
Here, $\Lambda$ starts from the initial value $\Lambda_0$ and is lowered up to 
$\Lambda \ll \xi^{-1}$.  
Namely, at the final point, we are in the hydrodynamic regime.  
The eliminated components turn out to be
 included in parameters of the Langevin equation.
Thus, the parameters, like the transport coefficients, are renormalized.

In other words, we first construct the nonlinear Langevin equation to describe 
 nonlinear effects in the critical regime, and the nonlinear effects are included
in the liner transport coefficients in the hydrodynamic regime by the RG transformation.

\subsection{Contrast with the static RG }

Here, we first stress that 
the concept of the dynamic universality class is not so universal contrary to its name.
Then, the class of the QCD critical point 
may not be the same as of the liquid-gas critical point or the model H
\footnote{The model H \cite{kawasaki1, HALPERIN:1969zza} is the minimal-dynamic model for a critical point
that its relevant modes are given as the nonrelativistic-hydrodynamic modes. 
The liquid-gas critical point belongs to the dynamic universality class of the model H}, 
although it is conjectured by \cite{son, moore}.
To see this, let us contrast the difference
 between the static RG with the dynamic one.

An important point is that the respective infrared effective theories are different;
in the static case, 
the infrared effective theory is the thermodynamic potential (or so-called Landau free energy). 
Then, its function form about order parameters are determined only by 
the space dimension and the symmetry among the order parameters but
not by microscopic details .
Thus, the concept of the universality class makes sense for the static case.
In contrast, for the dynamic RG,
the infrared effective theory is the nonlinear Langevin equation.

Here, the important difference arises; 
 the relevant variables for the Langevin equation is not only the order parameters 
but also conserved densities and NG modes, 
and its nonlinear couplings can not be determined by only the symmetry in general.
Consequently,  the dynamic universality class is not so universal compared to the static one.
Specifically, the nonlinear couplings, namely, the streaming terms $v_j(A)$, 
are generally given by the Poisson brackets (commutation relations)
 among the slow variables in the classical (quantum) system;
\begin{eqnarray}
v_j(A)=\sum\limits_{k}\bl[ Q_{j k}(A)\frac{\delta H}{\delta A_k}
-\beta^{-1}\frac{\delta }{\delta A_k}Q_{j k}(A) \br] ,
\label{eq: stream}
\end{eqnarray}
where
\begin{equation}
Q_{j k}(A)= \la \{ A_j, A_k \}_{\rm PB} ; A \ra \; \mbox{or}\;  \la [ A_j, A_k ]/(i \hbar )  ; A\ra. 
\end{equation} 
The above expression, Eq.(\ref{eq: stream}), is derived from Eq.(\ref{eq: vja}), see \cite{mazenko} for the derivation. 
The important point is that the Poisson-bracket relations depend on the microscopic expressions of the variables. 
This fact leads to an important consequence that the dynamic universality class of the QCD critical point 
may not be the same as  of the liquid-gas critical point or the model H.
Actually, in the model H, the Poisson-bracket relations are calculated with
 the non-relativistic relations \cite{kawasaki1, mazenko}.

\section{Slow variables near  the QCD critical point}

Here, we give earlier studies on the slow variables near the QCD critical point.
The slow variables have been identified as the fluctuations of 
the conserved densities:
the baryon number, $n$, and the energy and momentum, $T^{\mu \nu}$.

First, let us consider slow variables for chiral limit 
although we are interested in the  finite quark mass case.
For the chiral limit, the slow variables are $\sigma$, $\pi$,
and the conserved densities.
Here, $\sigma$ is the order parameter about the chiral phase transition and
$\pi$ is the NG mode for the spontaneous chiral-symmetry breaking.
Thus, the above quantities are slow.
The critical dynamics in this case is studied in \cite{Ohnishi:2004eb, Antoniou:2007tr}.

In contrast to the chiral limit, for finite quark mass,
the chiral symmetry is explicitly broken.
Thus, the pion has mass and is fast.
Moreover, by the explicit  symmetry breaking, $\sigma$ mixes to 
the conserved densities.
H. Fujii and M. Ohtani showed that, by this mixing, $\sigma $ also becomes massive, and
the flat direction of the thermodynamic potential is 
a linear combination of $\sigma$ and $n$ \cite{fujii}.

\begin{figure}[!t]
    \begin{minipage}{1.0\hsize}
     \centering
     \includegraphics[width=\hsize]{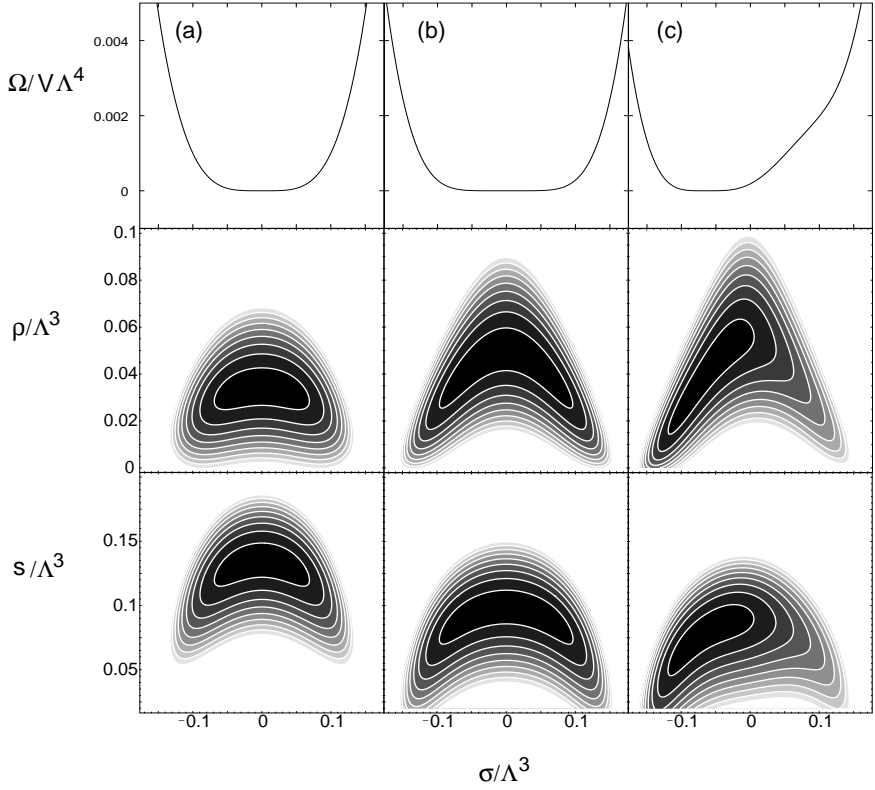}
       \caption{The thermodynamic potential in $(\sigma, \rho)$ and $(\sigma, s)$ planes.
                    Here, $\Omega$ is the thermodynamic potential, $\rho$ the baryon-number density.
                    and $s$ the entropy density.
                    For (a), the potential flats along the $\sigma$ direction 
                    whereas, for (c),  the potential flats along 
                    the linear combination of $\sigma$ and, $\rho$ or $s$. 
                    This figure is adapted from \cite{fujii}. }
     \label{fig: naname}
    \end{minipage}
\end{figure}

Figure \ref{fig: naname}, which is based on NJL model and adapted from \cite{fujii},
 shows the thermodynamic potential in $(\sigma, n)$ and $(\sigma, s)$ planes,
where, $s$ is the entropy density.
Here, (a) and (c) respectively denote the potentials for the chiral limits and for the finite quark mass.
Although (b) denotes for a tricritical point, we do not treat the point, see \cite{fujii} for the detail. 
We see that, for (a), the potential flats along the $\sigma$ direction
whereas, for (c),  the potential flats along the linear combination of $\sigma$ and, $n$ or $s$.
We note that, in this study, the quark mass is only a few MeV.  

Moreover, D.T. Son and M.A. Stephanov showed, by a linear Langevin equation, 
that the long-time behavior of the linear-combination mode
is determined only by the conserved densities \cite{son}.
Thus, $\sigma$ mode  just traces the conserved densities
and is unimportant.

Now,  we briefly give the study, \cite{son}.
First, to connect our Langevin equation, Eq.(\ref{eq: llangevin}),
to that in \cite{son}, we slightly rewrite Eq.(\ref{eq: llangevin}) as 
\begin{equation}
\frac{\partial}{\partial t} A_{j}(t) =
\sum\limits_k i\Omega_{j k}A_{k} (t)-\sum\limits_{k }L_{j k} \frac{\delta (\beta H(A) )}{\delta A_k}
+f_j(t). \label{eq: llangevin2}
\end{equation}
Here, we introduced the thermodynamic potential as the Gaussian form:
\begin{equation}
\beta H(A) = \int d\bfr \frac{1}{2}\bl[ \sum\limits_{i j} A_i \chi^{-1}_{i j} A_j \br] .\label{eq: betah} 
\end{equation}
If we substitute this potential in Eq.(\ref{eq: llangevin2}), 
we get back  Eq.(\ref{eq: llangevin}).
In the following, we consider Eqs.(\ref{eq: llangevin2}) and (\ref{eq: betah}).

For simplicity, we now neglect the energy and momentum densities
and consider the mixing only between the fluctuations of $\sigma $ and $n$.
Namely, our slow variables are 
\begin{equation}
\{ A_j \} = \{ \delta \sigma , \delta n\}
\end{equation}
In this case, the potential is\footnote
{In \cite{son}, derivative terms are included, but these are needless. }
\begin{equation}
\beta H(\delta \sigma, \delta n)= \int d \bfr \bl[ \frac{A}{2} 
(\delta\sigma )^2 + B \delta\sigma \delta n + \frac{C}{2} (\delta n)^2  \br],
\end{equation}
where $A, B$ and $C$ is related to inverses of the susceptibility.
The important point is that the second term explicitly break the chiral symmetry:
$\sigma \rightarrow -\sigma$.
By this term, we take into account the quark mass.

Then, we have the linear Langevin equation for $\sigma$ and $n$
in the Fourier space:
\begin{eqnarray}
\frac{\partial \delta \sigma (\bfk )}{\partial t} 
   &&= - L_{\sigma \sigma}(\bfk ) \frac{\delta (\beta H )}{\delta \sigma} 
       -L_{\sigma n}(\bfk )\frac{\delta (\beta H )}{\delta \sigma} + f_{\sigma}, \label{eq: sonsigma}\\
\frac{\partial \delta n (\bfk )}{\partial t} 
   &&= - L_{n n}(\bfk ) \frac{\delta (\beta H )}{\delta n} 
       -L_{n \sigma}(\bfk )\frac{\delta (\beta H )}{\delta \sigma} +f_{n}. \label{eq: sonn}
\end{eqnarray}
Here, time-reversible terms are absent by Eq.(\ref{eq: fmatrix2}) and the time-reversal symmetry.
In general, those terms vanish
if we have only variables whose time-reversal properties are even.

Now, let us consider the transport coefficients.
First, from the Onsager's reciprocal relation\cite{onuki}, 
we have
\begin{equation}
L_{\sigma n}(\bfk ) = L_{n \sigma}(\bfk ). 
\end{equation} 
Furthermore, we now expand the wavenumber dependence 
\begin{eqnarray}
L_{\sigma \sigma}(\bfk ) &&\sim \Gamma  + O(k^2), \\
L_{\sigma n}(\bfk ) &&\sim \tilde{\lambda} k^2 +O(k^4), \\
L_{n n} (\bfk) && \sim \lambda k^2 + O(k^4), 
\end{eqnarray}
because we are interested in the low-wavenumber region.
Here, $\Gamma, \tilde{\lambda}$ and $\lambda$ are wavenumber-independent constants.
The important point is that the expansions of $L_{n \sigma}$ and $L_{n n}$ start at the order $k^2$.
The reason is that
the baryon number density $n$ is conserved, 
and thus its time-change rate must vanish in the limit $\bfk \rightarrow 0$,
see Eqs.(\ref{eq: continuity}) and (\ref{eq: sonn}). 

Then, we finally arrive the linear Langevin equation at the leading order in $k$:
\begin{eqnarray}
\frac{\partial \delta \sigma (\bfk )}{\partial t} 
   &&= -\Gamma A \delta \sigma - \Gamma B \delta n  + f_{\sigma}, \\
\frac{\partial \delta n (\bfk )}{\partial t} 
   &&= -( \tilde{\lambda} A+\lambda B )k^2 \delta \sigma -( \tilde{\lambda} B+\lambda C )k^2 \delta n +f_{n}. 
\end{eqnarray}
From the above equations, we can obtain its eigen modes and dispersion relations.

Now, we leave the derivation to \cite{son} and give only the results.
The eigen modes are the following two modes:
\begin{eqnarray}
-B \delta \sigma + A \delta n \mbox{    and    } \delta \sigma .
\end{eqnarray}  
We see that these are the linear combination of $\sigma$ and $n$, and only $\sigma$.
These dispersion relations are
\begin{eqnarray}
\omega_{\sigma + n} (k) &&\sim -i D k^2, \\
\omega_{\sigma} (k) &&\sim  -i\Gamma A  + O(k^2),
\end{eqnarray}
where we  introduced the diffusion constant as 
\begin{equation}
D = \lambda C -\lambda \frac{B^2}{A}.
\end{equation}
We see that the linear-combination mode is slow, whereas the sigma mode has the gap 
in its dispersion, and thus it is fast.
Namely, the slow dynamics of  the linear-combination mode is determined only 
by the density fluctuation $\delta n$, whereas $\delta \sigma$ just traces $\delta n$.
Then, we have the only one slow mode. 

Now, we also write the spectrum (time correlation in Fourier space) of them:
\begin{eqnarray}
\la (\delta \sigma )^2 \ra &&=\frac{2 T \Gamma}{\omega^2 + \Gamma^2 }
                                                     +\frac{2 T\lambda k^2}{ \omega^2 + \lambda^2 k^4}, \\
\la (\delta n )^2 \ra &&=\frac{2 T\lambda k^2}{ \omega^2 + \lambda^2 k^4} \label{eq: son},
\end{eqnarray}
for a later comparison.

In this study, the coupling with the energy and momentum is not taken into account.
In the next chapter, we shall consider the couplings.
As a result, we shall find that actual slow modes is three:
thermal, viscous and sound modes.

\chapter{Linear dynamics of the hydrodynamic modes by relativistic hydrodynamics}
\label{sec:sdaqcp}

In this chapter, we study the slow dynamics 
near the QCD critical point in the hydrodynamic regime.

The slow variables for the QCD critical point is identified as 
the fluctuations of the conserved densities, as shown in the previous section.
Thus, we have the slow variables:
\begin{equation} 
\{ A_j \} =\{ \delta n, \delta e=(\delta T^{0 0}), \delta J^i=(\delta T^{0 i}) \}.
\end{equation}

In a straightforward way, 
we must construct the linear Langevin equation for them
from Eqs.(\ref{eq: llangevin}) and (\ref{eq: fmatrix2}).
However, we here develop the linear equation 
by linearizing relativistic hydrodynamic equation.
We note that the slow dynamics of the conserved densities is
basically given by the hydrodynamics. 
Then, the resulting linear equation is equivalent to that by 
the straightforward derivation. 

As relativistic hydrodynamic equations, we use the Landau equation
and the Israel-Stewart equation.
The Landau equation is believed to be an acausal.
Namely, a propagation speed of an information in the equation 
is considered to be faster than the light speed.  
In contrast, the Israel-Stewart equation has relaxation times,
and the causality problem is formally resolved.

However, we shall show that  the Landau equation has no problem to describe 
slowly varying fluctuations.
Furthermore, we shall find that the Israel-Stewart equation gives the same result 
the Landau equation gives in the long-wavelength region.    
 
The relativistic hydrodynamic equation is given by 
the following conservation laws
\begin{eqnarray}
\partial_{\mu}N^{\mu}=0, \label{eq: n}\\
\partial_{\mu}T^{\mu\nu}=0, \label{eq: T}
\end{eqnarray}
where $N^{\nu}$ and $T^{\mu \nu}$ 
are the baryon-number current and the energy-momentum tensor, respectively.
Those are given as
\begin{eqnarray}
N^\mu &=& n u^\mu+\nu^\mu, \\
T^{\mu \nu}&=&h u^{\mu}u^{\nu}-Pg^{\mu\nu}+\tau^{\mu\nu},  
\end{eqnarray}
where $h=e+P$ is the enthalpy density with $e$ 
and $P$ being the energy density and the pressure.
Also, $u^{\mu}=(\gamma, \gamma \bfv)$ are the fluid four velocity,
with $\gamma$ being the Lorentz factor, and
the dissipative terms, $\nu^{\mu}$ and $\tau^{\mu  \nu}$.
The dissipative terms differ among considered equations.

\section{For Landau equation }

For the Landau equation, the dissipative terms are  
\begin{eqnarray}
\nu^{\mu} =&& \lambda \bl( \frac{n T}{h} \br)^2 \partial_{\perp}^\mu (\beta \mu ), \label{eq: d1} \\
\tau^{\mu \nu}=&&\eta \bl[ \partial^{\mu}_{\perp}u^{\nu}
               +\partial^{\nu}_{\perp}u^{\mu}
               -\frac{2}{3}\Delta^{\mu\nu}(\partial_{\perp}{\cdot}u)\br]
               +\zeta\Delta^{\mu\nu}(\partial_{\perp}{\cdot}u), \label{eq: d2}                         
\end{eqnarray} 
where $\lambda$, $\eta$ and $\zeta$ are the thermal conductivity, the  share and 
 bulk viscosities, respectively.
 $\Delta^{\mu\nu} \equiv g^{\mu \nu}-u^{\mu} u^{\nu}$ is the projection onto the space-like vector 
and $\partial^{\mu}_{\perp} \equiv \Delta^{\mu \nu}\partial_{\nu}$ is  the space-like derivative.

Now, we linearize the Landau equation about fluctuations from the equilibrium values.
Let us write 
 $n(x)=n_c+\delta n(x)$, 
$e(x)=e_c + \delta e (x)$, $P(x)=P_c+\delta P(x)$, 
$(\beta \mu )(x)=(\beta\mu)_c + \delta (\beta\mu) (x)$, and $u^\mu (x)=u^\mu_c + \delta u^\mu (x)$.
Here, the symbols with a prefix $\delta$ denote the fluctuations.
The equilibrium values are denoted by a suffix $c$.
Hereafter, variables with the suffix and the prefix respectively denote the
equilibrium values and fluctuations.   

For simplicity, let us choose the rest frame as the reference frame:
$u_c^\mu=(1,\bf{0})$.
Then, by the relation $u_c^{\mu}\delta u_{\mu}=0$,
we have the fluid-velocity fluctuation as
\begin{equation}
\delta u^{\mu }= (0, \delta \bfv ).
\end{equation}  
We also note that the fluid-velocity fluctuation is related to the momentum density as
\begin{equation}
\delta \bfJ = h_c \delta \bfv
\end{equation}

Then,  Landau equation, Eqs.(\ref{eq: n})-(\ref{eq: d2}), are linearized as
\begin{eqnarray}
\frac{\partial \delta n}{\partial t}=&&-n_c \nabla \cdot \delta \bfv 
  +\lambda_0 \bl( \frac{n_c T_c}{h_c} \br)^2 \nabla^2 \delta ( \beta\mu ), \label{eq: ln}\\
\frac{\partial \delta e}{\partial t}=&& - h_c\nabla \cdot \delta \bfv , \label{eq: le}\\
\frac{\partial \delta\bfJ}{\partial t}=&& -\nabla(\delta P)
  +\bl( \zeta_0+\frac{1}{3}\eta_0 \br) \nabla(\nabla\cdot\delta\bfv )+\eta_0\nabla^{2}\delta\bfv .   \label{eq: lj}
\end{eqnarray}
This is the  linear equation of motion for our slow variables
\footnote{We can show that Eqs.(\ref{eq: ln})-(\ref{eq: lj}) are identical to 
 the linear Langevin equations derived from Eqs.(\ref{eq: llangevin}) and (\ref{eq: fmatrix2}),
 straightforwardly \cite{hidaka}.}.
This equation includes the couplings among the baryon number and the energy-momentum.
In the following, we study effects of the couplings on the density fluctuation $\delta n$.
Namely, we shall derive a spectral function of the density fluctuation from  Eqs.(\ref{eq: ln})-(\ref{eq: lj}).

Now, we have five equations for seven unknown quantities, $\delta n$, $\delta e$,
$\delta \bfJ$, $\delta P$, and $\delta (\beta \mu ) $.
To solve these equations, let us expand the thermodynamic quantities, 
 $\delta e$, $\delta P$, and $\delta (\beta \mu ) $, with the density and temperature fluctuations:
\begin{eqnarray}
\delta e =&&\bl( \frac{\partial  e}{\partial n}\br)_T \delta n 
             + \bl( \frac{\partial  e}{\partial T}\br)_n \delta T, \\
\delta P =&&\bl( \frac{\partial  P}{\partial n}\br)_T \delta n 
             + \bl( \frac{\partial  P}{\partial T}\br)_n \delta T,  \\         
\delta (\beta \mu) =&& \bl( \frac{\partial  (\beta \mu )}{\partial n}\br)_T \delta n 
             + \bl( \frac{\partial  (\beta \mu )}{\partial T}\br)_n \delta T.
\end{eqnarray} 
The merit of the set $(\delta n , \delta T)$ is that 
their equal-time correlation is orthogonal 
\begin{equation}
 \la \delta n (\bfk, t )\delta T(-\bfk, t) \ra =0,
\end{equation}
for a grand canonical ensemble \cite{onuki, mazenko}.

In terms of $(\delta n, \delta T, \delta \bfv)$,
 the equations (\ref{eq: ln})-(\ref{eq: lj}) take the form
\begin{eqnarray}
\bl( \frac{\partial}{\partial t}&&-\lambda\frac{T_0 c_s^2 c_v}{h_c c_p}\nabla^2 \br) \delta n
 +n_0\nabla\cdot\delta\bfv \nonumber \\
 &&+\lambda\frac{n_0}{h_c}\bl( 1-\frac{c_s^2 \alpha_P T_0 c_v}{c_p}\br) \nabla^2 \delta T=0, \\
h_{c}\frac{\partial \delta\bfv }{\partial t}&&-\eta\nabla^{2}\delta\bfv 
 -\bl( \zeta+\frac{1}{3}\eta \br)\nabla(\nabla\cdot\delta\bfv ) \nonumber \\
 &&+\frac{h_c c_s^2 c_v}{n_0 c_p}\nabla \delta n
 +\frac{h_c c^2_s c_v\alpha_P}{c_p}\nabla \delta T=0,                        \\
\bl( -\frac{h_c c_s^2 c_v\alpha_P }{n_0 c_p }\frac{\partial }{\partial t}&&
  +\lambda \frac{c_s^2}{n_0\gamma } \nabla^2  \br) \delta n \nonumber \\
 &&+\bl[ \frac{n_0 c_v}{T_0} \frac{\partial }{\partial t}
  +\lambda \bl( \frac{c_s^2 c_v\alpha_P }{c_p }-\frac{1}{T_0}\br) \nabla^2 \br] \delta T=0,
\end{eqnarray}
where  
$c_v=T_0(\partial s / \partial T)_n$ and $c_p=T_0(\partial s / \partial T)_P$
 are the specific heats at constant volume and pressure, respectively, 
$c_s=(\partial P/\partial e)_s^{1/2}$ the sound velocity, 
$\alpha_P=-(1/n_c)(\partial n / \partial T)_P$  the thermal expansivity at constant pressure.
Here, we used some thermodynamic identities,
see \cite{Minami:2009hn} for detail.

Now, let us perform Fourier-Laplace transformation, like 
\begin{equation*}
\delta n(\bfk ,z)=\int_{-\infty}^{+\infty}d\bfr \int^{\infty}_{0}dt
\: \e ^{- z t-i\bfk \cdot\bfr } \delta n(\bfr ,t).
\end{equation*}
Then, we find
\begin{eqnarray}
\bl( z + k^2 \lambda\frac{T_c c_s^2 c_v}{h_c c_p} \br) && \delta n (\bfk ,z)
 +i n_c \bfk  \cdot \delta\bfv (\bfk ,z)  \nonumber \\
 &&+k^2 \lambda\frac{n_c}{h_c}\bl( \frac{c_s^2 c_v\alpha_P T_c}{c_p}-1\br) \delta T  
 =\delta n(\bfk ,t=0),          
\label{eq:fln}             \\
\bl( z h_c +k^2 \eta \br) \delta\bfv (\bfk ,z) &&
 +\bl( \zeta+\frac{1}{3}\eta \br) \bfk (\bfk  \cdot \delta\bfv (\bfk ,z)) 
 +i \bfk  \frac{h_c c_s^2 c_v}{n_c c_p}\delta n(\bfk ,z)   \nonumber \\
 &&+i\bfk  \frac{h_c c^2_s c_v\alpha_P}{c_P} \delta T(\bfk ,z)
 =h_c \delta \bfv (\bfk ,t=0),        
\label{eq:flm}             \\ 
-\bl( z\frac{h_c c_s^2 c_v\alpha_P }{n_c c_p } 
 +k^2 \lambda \frac{c_s^2 c_v}{n_c c_p }   \br) &&\delta n(\bfk ,z) 
 +\bl[ z \frac{n_c c_v}{T_c}  
 -k^2 \lambda \bl(\frac{c_s^2 c_v\alpha_P }{c_p }-\frac{1}{T_c}\br) \br] \delta T(\bfk ,z) \nonumber \\
 &&=-\frac{h_c c_s^2 c_v\alpha_P }{n_c c_p }\delta n(\bfk ,t=0)
  +\frac{n_c c_v}{T_c}\delta T(\bfk ,0).
\label{eq:fls}
\end{eqnarray}
Here, we note that the initial values, like $\delta n ( \bfk, t=0) $, arise from the time derivative terms,
because we performed the Laplace transformation about time.

It is convenient to divide the velocity into longitudinal and transverse components
\begin{eqnarray}
\delta v_{\parallel}(\bfk, z) &&\equiv \hat{\bfk} \cdot \delta \bfv (\bfk, z) \\
\delta \bfv_{\perp }(\bfk ,z) && \equiv  \delta \bfv (\bfk, z) - \hat{\bfk} \delta v_{\parallel}(\bfk, z)
\end{eqnarray}
The transverse component of Eqs.(\ref{eq:fln})-(\ref{eq:fls}) reads 
\begin{equation}
(z h_c+k^2 \eta )\delta\bfv_{\perp }(\bfk ,z)
 =h_c \delta\bfv_{\perp } (\bfk ,t=0).
 \label{eq:transverse}
\end{equation}

Now, let us first study the transverse component. 
The solution is given by  
\begin{equation}
\delta \bfv_{\perp }(\bfk, z) =\frac{\delta \bfv_{\perp }(\bfk,0)}{z+(\eta /h_c)k^2 }. \label{eq: dvp} 
\end{equation}
Performing the inverse Laplace transformation 
\begin{equation}
\delta \bfv_{\perp }(\bfk ,t)=\frac{1}{2\pi i} \int ^{\delta+i\infty }_{\delta-i\infty}dz
\e ^{z t} \delta \bfv_{\perp }(\bfk ,z),
\end{equation}
we have 
\begin{equation}
\delta \bfv_{\perp }(\bfk ,t) = e^{- (\eta/h_c)k^2 t} \delta \bfv_{\perp }(\bfk , 0). 
\end{equation}
We see that the transverse component of the momentum diffuses, without propagation.
This modes is called the viscous diffusion mode.

Furthermore, let us derive spectral function of $\delta \bfv_{\perp }$.
Performing Fourier transformation about time $t$,
we obtain
\begin{equation}
\delta \bfv_{\perp }(\bfk, \omega ) =\frac{(\eta /h_c) k^2}{ \omega^2 + (\eta/h_c )^2 k^4}
                                                   \delta \bfv_{\perp} (\bfk, 0). \label{eq: vp}
\end{equation}
Multiplying Eq.(\ref{eq: vp}) by the initial value $\delta \bfv_{\perp} (\bfk, 0)$
and taking the statistical average,
we have the spectral function
\begin{eqnarray}
S_{\perp \perp}(\bfk, \omega ) &&\equiv \la \delta \bfv_{\perp }(\bfk, \omega ) \delta \bfv_{\perp }(\bfk, 0 ) \ra ,
\nonumber \\
&&= \frac{(\eta /h_c) k^2}{ \omega^2 + (\eta/h_c )^2 k^4} \la (\delta \bfv_{\perp} (\bfk ,0) )^2\ra . \label{eq: svp}
\end{eqnarray} 
Here, we note that $ \la (\delta \bfv_{\perp} (\bfk ,0) )^2\ra $ does not have information on the time evolution.

Now, Let us return to the longitudinal component. 
The longitudinal component of Eqs.(\ref{eq:fln})-(\ref{eq:fls}) can be written as  
 the following matrix form
\begin{equation}
A  
  \begin{pmatrix}
  \delta n(\bfk ,z) \\
  \delta v_{\parallel}(\bfk ,z) \\
  \delta T(\bfk ,z)
  \end{pmatrix}
  =  
  \begin{pmatrix}
  \delta n(\bfk ,0) \\
  \delta v_{\parallel}(\bfk ,0) \\
  -\frac{\alpha_{P}c_{s}^{2}c_v}{n_c c_p}\delta n(\bfk ,0)
  +\frac{n_c c_v}{T_c h_c}\delta T(\bfk ,0)
  \end{pmatrix},
\label{eq:matrix}
\end{equation}
where the matrix A is 
\begin{equation}
A= 
  \begin{pmatrix}
  z + k^2 \lambda \frac{T_c  c_{s}^{2}c_v}{h_c c_p}  & i k n_c & 
  -k^2 \lambda \frac{n_c}{h_c}(1-\frac{\alpha_P c_{s}^{2} c_v T_c}{c_p})  \\
 i k \frac{c_{s}^{2}c_v}{n_{0}c_p} 
  & z+\nu_{l}k^{2} 
  & i k\frac{\alpha_{P} c_{s}^{2}c_v}{c_p}\\
 \frac{n_c c_v}{h_c T_c}
  [-z\frac{h_c T_c\alpha_{P}c_{s}^{2}}{n_c^2
    c_p}-k^{2} D_{\rm t} c_{s}^{2}\frac{T_c}{n_c}] 
  & 0
  & \frac{n_c c_v}{h_c T_c}
    [z+k^{2}\frac{c_p}{c_v} D_{\rm t} (1-\frac{\alpha_{P}c_{s}^{2}T_c c_v}{c_p})]
\end{pmatrix}.
\end{equation}
Here, we introduced the longitudinal kinetic-viscosity $\nu_l$, and
the thermal diffusion constant $D_{\rm t}$, 
\begin{eqnarray}  
\nu_l &&=\bl( \zeta+\frac{4}{3}\eta \br) /h_c ,\\
D_{\rm t} &&=\frac{\lambda}{n_c c_p}.
\label{eq:long-kin}
\end{eqnarray}
Multiplying the inverse $A^{-1}$ from the left in Eq.(\ref{eq:matrix}),
we obtain the Fourier-Laplace coefficient of the density fluctuation  
\begin{eqnarray}
 \delta n(\bfk ,z)=\bl[ (A^{-1})_{1 1} 
 -\frac{\alpha_{P}c_{s}^{2} c_v}{n_{c}c_p}(A^{-1})_{1 3} \br] \delta n(\bfk ,0)
+(A^{-1})_{1 2} \delta v_{\parallel}(\bfk ,0)    \nonumber                   \\
+\frac{n_{c}\tilde{c}_{n}}{T_{c}h_c} (A^{-1})_{1 3} \delta T(\bfk ,0). \label{eq:tilden}
\end{eqnarray}
Here, an important point is that $\delta n$ is orthogonal to $\delta T$ and $\delta \bfv$:
\begin{eqnarray}
\la \delta n(\bfk ,0) \delta T (\bfk ,0) \ra &&= 0, \\
\la \delta n(\bfk ,0)  \delta v_{\parallel} (\bfk ,0) \ra &&=0.
\end{eqnarray}
The second equation comes from the time-reversal invariance of the equilibrium state. 

Thus, by the similar procedure as in the transverse component,
we obtain the spectral function of the density fluctuation 
\begin{eqnarray}
S_{n n}(\bfk ,\omega ) &=& \la \delta n(\bfk ,\omega ) \delta n(\bfk ,t=0)\ra \nonumber \\
   &=& \la (\delta n(\bfk ,t=0))^2\ra \bl[ \;\bl( 1-\frac{c_v}{c_p} \br)
   \frac{2 D_{\rm t} k^{2}}{\omega^{2}+D^{2}k^{4}}
   \nonumber \\
   &+&\frac{c_v}{c_p}
   \bl( \frac{D_{\rm s} k^{2}}{(\omega -c_{s}k)^{2}+D_{\rm s}^{2}k^{4}}
   +\frac{D_{\rm s} k^{2}}{(\omega +c_{s}k)^{2}+D_{\rm s}^{2}k^{4}}\br) \; \br],
   \label{eq:landau}
\end{eqnarray}
where we have introduced the sound diffusion constant, $D_{\rm s}$, as
\begin{equation}
D_{\rm s} =\frac{1}{2} \bl[ D_{\rm t} \bl( \frac{c_v}{c_p}-1\br) +\nu_{l}\br] 
+\frac{c_s^2 T_c}{2} \bl( \frac{\lambda}{h_c} -2 D_{\rm s} \alpha_P \br) .
\end{equation}
The detailed derivation is given in Appendix \ref{sec:dds}. 
We see that the spectral function has three peaks at frequencies $\omega=0$ and 
$\omega = \pm c_s k$:
The peak at $\omega=0$ corresponds to thermally induced density fluctuations.
This mode is  called the thermal diffusion mode.
The two side peaks at $\omega=\pm c_s k$ correspond to mechanically induced density 
fluctuation, i.e. sound waves.  
This mode is called the sound mode.
Roughly speaking, if we expand the density, $\delta n$, 
with the entropy density, $\delta s$, and the pressure, $\delta P$,
\begin{equation}
\delta n  = \bl( \frac{\partial n}{\partial s}\br)_{P} \delta s + \bl( \frac{\partial n}{\partial P}\br)_{s} \delta P,
\end{equation}  
we see that the first term corresponds to the thermal mode, while the second term the sound mode.

Now, let us  compare this result with that in the  non-relativistic case\cite{lp, mazenko};
\begin{eqnarray}
S^{\rm NR}_{n n}(\bfk ,\omega ) 
   &&= \la (\delta n(\bfk ,t=0))^2\ra \bl[ \; \bl( 1 - \frac{c_v}{c_p} \br)
   \frac{2 D_{\rm t} k^{2}}{\omega^{2}+D_{\rm t}^2 k^{4}}
   \nonumber \\
  &+&\frac{c_v}{c_p}
   \bl( \frac{D_{\rm s}^{\rm NR} k^{2}}{(\omega -c_{s}k)^{2}+
{D_{\rm s}^{\rm NR}}^{2}k^{4}}
   +\frac{D_{\rm s}^{\rm NR} k^{2}}
{(\omega +c_{s}k)^{2}+{D_{\rm s}^{\rm NR}}^{2}k^{4}}\br) \; \br],
   \label{eq:navier}
\end{eqnarray} 
where 
\begin{eqnarray}
D_{\rm s}^{\rm N R} &&=\frac{1}{2}\bl[ D \bl( \frac{c_p}{c_v}-1\br) +\nu_{l}^{\rm N R} \br],\\
\label{eq:gnr}
\nu_{l}^{\rm N R} &&=\bl( \zeta + \frac{4}{3}\eta \br) / \rho_c . \label{eq: nulnr}
\end{eqnarray}
We see that  relativistic effects 
appear only in the sound diffusion constant:
\beq
D_{\rm s}=D_{\rm s}^{\rm MR}+\delta D_{\rm s},
\eeq
where
\begin{eqnarray}
D_{\rm s}^{\rm MR} &&\equiv \frac{1}{2}
\bl[ D_{\rm t} \bl( \frac{c_p}{c_v}-1\br) +\nu_l \br] ,
\end{eqnarray}
and
\begin{equation}  
\delta D_{\rm s} \equiv \frac{c_s^2 T_c}{2} \bl( \frac{\lambda}{h_c} -2 D_{\rm s} \alpha_P \br) .
\end{equation}
First, the longitudinal kinetic viscosity is expressed in terms of the enthalpy
density $h_c$ in the relativistic case in place of the mass density $\rho_c$,
see Eqs.(\ref{eq:long-kin}) and (\ref{eq: nulnr}).
We call this modification the minimal-relativistic (MR) effect.

Next, the other is a genuine relativistic effect $\delta D_{\rm s}$
which is absent in the non-relativistic case.
This part comes from the dissipative term of Eq. (\ref{eq: ln}), 
which represent relativistic effects, and 
vanishes if we take the light speed $c \rightarrow \infty$.

To see the relativistic effects $\delta D_{s}$ quantitatively,
we now determine thermodynamic quantities by
the equation of state(EoS) of massless classical ideal gas, $e=3 P=3 n T$. 
Then, we have $c_p=4$, $c_v=3$, $\alpha_P = 1 / T_c$, $c_s = \sqrt{1/3}$ 
and the entropy density, $s_{c}=4 n_c -  \mu_c n_c  /T_c $.

\begin{figure}[!t]
 \begin{minipage}{1.0\hsize}
  \centering
   \includegraphics[width=70mm,angle=270]{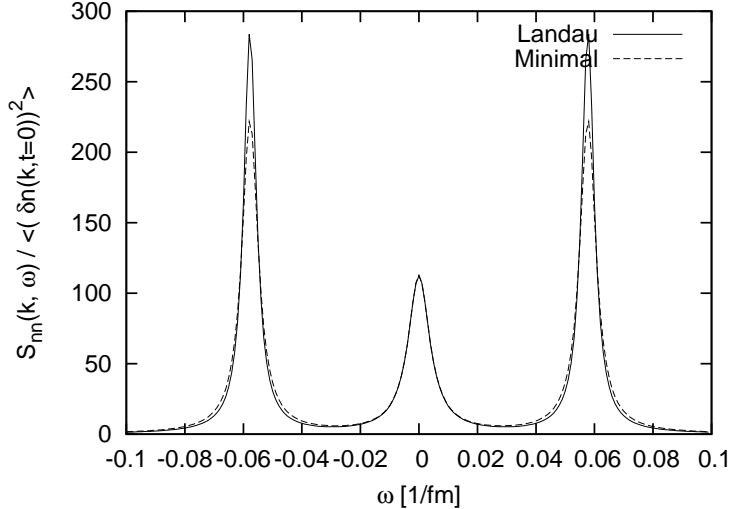}
  \caption{The spectral function for the Landau  and  
           minimal relativistic equation. The solid 
           and dashed lines respectively denote the Landau and minimal cases.
           The parameters are $k=0.1$[1/fm],\, $\mu_c=200$[MeV],\, 
           $T_c=200$[MeV],\, $\eta/(n_c s_c)=\zeta/(n_c s_c)=0.3$ and
           $\lambda T_c/(n_c s_c)=0.6$. 
           Relativistic effects does not appear in the thermal mode
           but enhance the sound modes.}
  \label{fig:landau}
 \end{minipage}
\end{figure}

Figure \ref{fig:landau} shows the spectral function, Eq.(\ref{eq:landau}), and 
 the minimal relativistic case with the above thermodynamic quantities.
The parameter set is given as the following; 
$k=0.1$[1/fm],\, $\mu_c=200$[MeV],\, 
$T_c=200$[MeV],\, $\eta / s_c =\zeta / s_c=0.3$ and $\lambda T_c/ s_c=0.6$.

As is expected,  Fig.\ref{fig:landau} shows that the sound mode
 is enhanced by the relativistic effects,
while the thermal mode is the same as in the minimal case.

Now, we also compare our results to Eq.(\ref{eq: son}) in which 
the coupling among $\delta n$, $\delta e$ and $\delta \bfJ$ is neglected.
The spectrum (\ref{eq: son}) has the only one slow mode, which is diffusive.
In contrast, our result Eq.(\ref{eq:landau}) has the two modes:
the thermal and sound modes.
In addition, we have the another slow mode in the transverse component Eq.(\ref{eq: svp}):
the viscous mode.

Namely, the actual slow modes is three.
In Sec.\ref{sec:taqcp}, 
we shall study the relative importance of them near  the QCD critical point.

\section{For Israel-Stewart equation}
\label{sec: is}

For the Israel-Stewart equation in the particle frame, 
the dissipative terms are 
\begin{equation}
\tau^{\mu \nu}=-\Pi\Delta^{\mu \nu}+q^\mu u^\nu +q^\nu u^\mu +\pi^{\mu \nu}, 
\end{equation}
and
$\nu^\mu=0$.
Here 
\beq
\Pi &=& 
-\zeta(\partial_\mu u^\mu +\beta_0 u^\mu\partial_\mu \Pi -\alpha_0 \partial_\mu q^\mu),\\ 
q^\mu& =&\lambda T \Delta^{\mu \nu}\bl(\frac{1}{T}\partial_\nu T
       -u^\rho\partial_\rho u_\nu 
       -\beta_1 u^\rho\partial_\rho q_\nu
       -\alpha_0 \partial_\nu \Pi +\alpha_1 \partial_\rho \pi^\rho_\nu  \br), \\
\pi^{\mu \nu}&=& 2\eta \Delta^{\mu \nu \rho \sigma}( \partial_\rho u_\sigma
              -\beta_2 u^\tau \partial_\tau \pi_{\mu \rho}
              -\alpha_1 \partial_\rho q_\sigma  )  ,
\eeq
with 
$u^\mu q_\mu=0,$ \, 
$\pi^{\mu \nu}=\pi^{\nu \mu}$,\, 
$u^\mu \pi_{\mu \nu}=0$ and 
$\pi^\mu_\mu=0$.
Here, $\beta_0 $, $\beta_1 $ and $\beta_2 $ are the relaxation time of   
the bulk viscous, the heat flux and the shear viscous, respectively.
$\alpha_0$ ($\alpha_1$) is the coupling of the bulk viscose and the heat flux 
(the shear viscose and the heat flux).
 $\Delta^{\mu \nu \rho \sigma}$ is the projector defined by   
\begin{equation}
\Delta^{\mu \nu \rho \sigma} =\frac{1}{2}\bl[ \Delta^{\mu \rho}\Delta^{\nu \sigma}
                      +\Delta^{\mu \sigma} \Delta^{\nu \rho} 
                      -\frac{2}{3}\Delta^{\mu \nu} \Delta^{\rho \sigma} \br], 
\end{equation} 
Here, an important point is that
the relaxation time corresponds to a correlation time of the memory function, 
Eq.(\ref{eq: memory}).
Namely, the Israel-Stewart equation has the memory effect, but such effect is irrelevant for the slow dynamics,
as we shall see in the following. 

Applying the similar procedure as in the Landau equation,
we have the spectral function of the density fluctuation:  
\begin{eqnarray}
 \frac{S_{n n}(\bfk ,\omega )}{\la (\delta n(\bfk ,t=0))^2\ra }=
  &&\bl( 1-\frac{c_v}{c_p}\br) \frac{2 D k^{2}}{\omega^{2}+D^{2}k^{4}} 
   +\frac{c_v}{c_p} \bl[ \frac{ D_s k^{2}}{(\omega -c_{s}k)^{2}+D_s^{2}k^{4}} \nonumber \\
     &&+\frac{D_s k^{2}}{(\omega +c_{s}k)^{2}+D_s^{2}k^{4}} \br]      
     +O(k^2) \times \bl[ \frac{2  /\beta_0\zeta}{\omega^{2}+1/(\beta_0\zeta )^2}  \nonumber \\  
       &&+\frac{1  /\beta_2\eta}{\omega^{2}+1/(2\beta_2\eta)^2}
       +\frac{2 h_c /[(\beta_1 h_c-1)\lambda T_0]}
       {\omega^{2}+h_c^2 /[(\beta_1 h_c-1)\lambda T_c]^2} \br].
 \label{eq:is}
\end{eqnarray}
We leave the detailed derivation to \cite{Minami:2009hn}.
Here, we assumed the relaxation time satisfies the inequality, $\beta_1 > 1/h_c$.
If the relaxation time does not satisfy the inequality, 
we have a pathological behavior; the spectrum becomes negative.
The Israel-Stewart equation in particle frame takes over the pathological 
behavior of the Eckart equation, in which the fluctuation does not relax;
see the detailed discussion in \cite{Minami:2009hn}.

Now, the spectral function apparently has six peaks including the conventional
three peaks,
but the new  three Lorentzian functions should vanish in the long-wavelength limit 
$k \rightarrow 0$,
because the strength of these is of the second order in $k$.
Therefore, the Israel-Stewart equation gives
 the same result for the hydrodynamic spectrum as the Landau equation does
in the long-wavelength limit.
Namely, the relaxation times does not affect the result in the long-wavelength region.
This result implies that the causality problem occurs only in the short-wavelength region.
We note that such illegal component would be ruled out by the cutoff, as mentioned in Sec.\ref{sec:osv}    

\section{Tendency around the QCD critical point}
\label{sec:taqcp}

Here, we include a part of information on the critical point in the hydrodynamic spectrum
by static scaling laws.
By this study, we shall find the tendency of the slow variables around the QCD critical point;
namely, which fluctuations are enhanced  near the critical point.
We note that not all fluctuations are enhanced near the critical point.

Now, we use the static scaling laws:
\begin{eqnarray}
c_v &&\sim \xi^{\alpha / \nu}, \\
c_p &&\sim \xi^{\gamma / \nu},
\end{eqnarray}
where $\xi$ is the correlation length, which diverges at the critical point.
$\alpha, \nu$ and $\gamma$ are usual static critical exponents.
The QCD critical point belongs to the static universality class of the 3d Ising model, $Z(2)$.
Then,   we have the critical exponents
\begin{eqnarray}
\alpha \sim 0.1, \\
\nu \sim 0.6, \\
\gamma \sim 1.2,
\end{eqnarray}
and the critical behaviors of the specific heats are
\begin{eqnarray}
c_v &&\sim \xi^{0.2}, \\
c_p &&\sim \xi^2.
\end{eqnarray}

Then, we have the critical behavior of the spectrum, Eq.(\ref{eq:landau}),
\begin{eqnarray}
S_{n n}(\bfk ,\omega )  &/& \la (\delta n(\bfk ,t=0))^2 \ra
   =  \bl[ \;\bl( 1-\frac{c_v}{c_p} \br)
   \frac{2 D_{\rm t} k^{2}}{\omega^{2}+D_{\rm t}^{2}k^{4}}
   \nonumber \\
   &+&\frac{c_v}{c_p}
   \bl( \frac{D_{\rm s} k^{2}}{(\omega -c_{s}k)^{2}+D_{\rm s}^{2}k^{4}}
   +\frac{D_{\rm s} k^{2}}{(\omega +c_{s}k)^{2}+D_{\rm s}^{2}k^{4}} \br) \; \br], \\
   &\sim & \frac{2 D_{\rm t} k^2}{\omega^{2}+D_{\rm t}^2 k^4}, \label{eq: csnn} 
\end{eqnarray} 
because $c_v/c_p $ behaves as $ \xi^{-1.8}$ and vanishes at the critical point.
We see that the sound mode disappears, whereas the thermal mode survives. 

Moreover, the thermal diffusion constant behaves as
\begin{equation}
D_{\rm t} = \frac{\lambda }{n_c c_p} \sim \xi^{-2} \rightarrow 0 \label{eq: Dt}.
\end{equation}
Then, the surviving thermal mode is enhanced near the critical point.
This is the critical-slowing down.
\begin{figure}[!t]
  \centering
   \includegraphics[width=70mm, angle=270]{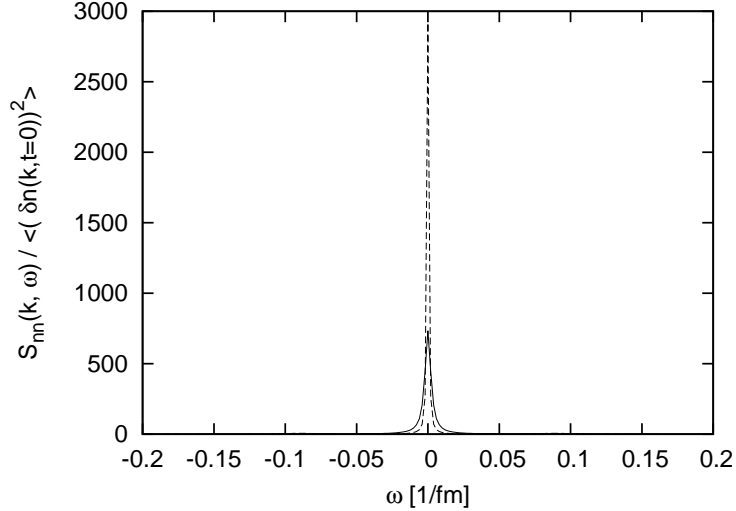}
  \caption{ A schematic figure for the density spectrum near the critical point.
                The sound mode disappears, whereas the thermal mode is enhanced. }
  \label{fig:t1}
\end{figure}

Meanwhile, the transverse spectrum, Eq.(\ref{eq: svp}), has no critical behavior.
Then, the viscous mode is not enhanced nor suppressed near the critical point.

Now, we see that relative importance of the three hydrodynamic modes.
\begin{enumerate}
\item The thermal mode is the most relevant mode
         because it  survives and is largely enhanced near the critical point.
\item The viscous mode is the second relevant mode.
         The mode is not enhanced nor suppressed.
\item The sound mode is suppressed and thus not so important.
         For the minimal critical dynamics, we can neglect the sound mode.
         However, for the renormalization of the bulk viscosity in the next chapter,
         we must consider the sound mode.
         An effect on the thermal mode by the sound mode is negligible, 
         whereas a counter effect is large.     
\end{enumerate}

From the above, we can know the tendency of the slow variables near the critical point.
We first see that the density fluctuation, $\delta n$, is enhanced by the thermal mode, 
see Eqs.(\ref{eq: csnn}) and (\ref{eq: Dt}), and Fig. \ref{fig:t1}.
In addition, the energy density is also enhanced, because it  also couples to the thermal mode.

In contrast, we see that 
 the momentum $\delta \bfJ$ is not enhanced\footnote{
We note that the momentum density is proportional to the fluid-velocity fluctuation:
$\delta \bfJ = h_c \delta \bfv$. },
because the transverse and longitudinal components of that 
couple to only the viscous and sound modes, respectively.
These modes are not enhanced near the critical point.

Here, the critical behavior of the transport coefficients is not taken into account.
However, we note that the similar analysis on the liquid-gas critical point,
as in this section, qualitatively gives a good description\cite{mazenko}.

\chapter{Nonlinear dynamics of the hydrodynamic modes by the dynamic RG}
\label{sec:drgqcd}

Here, we study the nonlinear dynamics in the critical regime by the dynamics RG.
We note that the nonlinear effects, which is not included in the usual hydrodynamics, arise in the critical regime.
Thus, the analysis in the previous chapter is valid only in the hydrodynamic regime. 

In this chapter, we first construct the nonlinear Langevin equation
to describe the nonlinear dynamics in the critical regime.
Next, we include the nonlinear interaction in the transport coefficients by the dynamic RG.
Thus, the transport coefficients are renormalized and diverge at the critical point.  

\section{The nonlinear Langevin equation for the QCD critical point}

Now, let us construct the nonlinear Langevin equation.
An important point on the construction is that 
$\delta n$ and  $\delta e$ are enhanced, 
while $\delta \bfJ$ is not near the QCD critical point, as mentioned in Sec.\ref{sec:taqcp}.
Then, we can neglect the nonlinearity of the momentum
in the following construction.

\subsection{Thermodynamic potential for the slow variables}

First, we construct the thermodynamic potential 
$H(\delta n,\delta e, \delta \bfJ)$.

Because the momentum density fluctuation is not enhanced near the QCD critical point,  
we now assume the potential for the momentum as the Gaussian form.
Then, we have
$H(\delta n,\delta e, \bfJ)=H_{n e}(\delta n,\delta e)+H_{J}(\delta \bfJ)$,
with
\begin{equation}
H_{J}(\delta \bfJ) \equiv \frac{1}{2 h_c}\delta \bfJ^2.
\end{equation} 
In contrast to $\delta \bfJ$,
$\delta n$ and $\delta e$
are enhanced near the QCD critical point, the thermodynamic
potential $H_{ n e}(\delta n,\delta e)$ should contain 
 higher order terms of them.

Now, the important point is that $H_{ n e}(\delta n, \delta e)$ is 
the quantity to determine the static property of the system
and the QCD critical point belongs to the same static universality class as the 3d Ising class, namely, $Z_2$.
Therefore, we may construct $H_{n e}(\delta n, \delta e)$ 
with the thermodynamic potential for the 3d Ising system \cite{glw}, 
which reads
\begin{eqnarray}
\beta H_{\rm Ising}(\psi, m) =&&\int d\bfr 
   \bl[ \frac{1}{2}r_0\psi^2+\frac{1}{2}K_0|\nabla \psi|^2
   +\frac{1}{4}u_0\psi^4  \nonumber \\ &&
   +\gamma_0 \psi^2 m +\frac{1}{2 C_0}m^2 -h\psi-\tau m\br] .\label{eq: glw}
\end{eqnarray}
Here, $\psi$  and $m$ are the spin density and the exchange energy density, respectively.
$r_0$, $K_0$, $u_0$, $\gamma_0$ and $C_0$ denote the static parameters,
while $h$ and $\tau$ the applied magnetic field and the reduced temperature,
respectively.
Then, we assume  the thermodynamic potential as
\begin{equation}
H(\delta n,\delta e, \delta \bfJ )=H_{\rm Ising}(\psi, m)
 + \frac{1}{2 h_c}\delta \bfJ^2, \label{eq: potential} 
\end{equation}
if we have the mapping between $(\psi,\, m)$ and
$(\delta n,\, \delta e)$.

The general mapping relation between a grand canonical ensemble in $Z_2$ and the 3d Ising system is
known in condensed matter physics \cite{onuki}, which are summarized as follows.
First,
we assume the following  linear relation between the deviations of the intensive variables
from  the critical points, 
\footnote{Recall that the static scaling laws
are expressed by the deviations of the intensive variables from those at the critical point.}
\begin{eqnarray}
\delta h=&&\alpha_1\delta (\mu/T) +\alpha_2 \delta T/T_c , \label{eq: mr1}\\
\delta \tau =&&\beta_1\delta (\mu/T) +\beta_2 \delta T/T_c, \label{eq: mr2}
\end{eqnarray}
where 
$\alpha_1$, $\alpha_2$, $\beta_1$ and $\beta_2$ are constants and
assumed to be regular at the critical point.
We note that $\alpha_1$, $\alpha_2$, $\beta_1$ and $\beta_2$ need not to be determined for
 the critical divergence,
because those have no singularities at the critical point. 
Although one could use 
Eqs. (\ref{eq: mr1}) and (\ref{eq: mr2}) for the mapping,
  it turns out to be inconvenient 
for  a Langevin equation.
To translate these relations to more convenient ones,
we assume the following relation \cite{onuki}:
\begin{equation}
\psi\delta h  + m \delta \tau 
=T_c^{-2}\delta T\delta e+\delta(\mu/T)\delta n, \label{eq: mr3}
\end{equation}
which is actually derived by considering 
a change of the microscopic distributions 
by small deviations of the intensive variables in both systems.
From the relations Eqs. (\ref{eq: mr1})-(\ref{eq: mr3}), we arrive at the
convenient mapping relation as follows,
\begin{eqnarray}
\delta n =&& \alpha_1\psi +\beta_1 m,
\label{eq: map1} \\
T_c^{-1}\delta e =&& \alpha_2\psi +\beta_2 m. 
\label{eq: map2}
\end{eqnarray}
With this mapping,
Eq. (\ref{eq: potential}) now gives the thermodynamic potential for the QCD critical point.
We note that we only map the static quantities,
 although the dynamic ones are studied. 
For later uses, we introduce fluctuations of the intensive variables as
\begin{eqnarray}
\delta T &&\equiv T_c^2 \frac{\delta(\beta H)}{\delta e}, \label{eq: tf}\\ 
\delta \bl( \frac{\mu}{T} \br) &&\equiv \frac{\delta(\beta H)}{\delta n} .\label{eq: muf}
\end{eqnarray}
This relation comes from the fact that, 
in the grand canonical distribution $P_{\rm gra} \propto \exp[(1/T) e + (\mu/T) n]$,
 $e$ and $n$ are respectively conjugate to $1/ T$ and $\mu /T$
\cite{onuki}.
We also introduce the fluid velocity fluctuation as in the non-relativistic case:
\begin{equation}
\delta \bfv \equiv  \frac{\delta H}{\delta \bfJ}. \label{eq: vf}
\end{equation}

We note that the static parameters in Eq. (\ref{eq: glw}) 
has the ultraviolet cutoff dependence in the region $\xi^{-1}<\Lambda$.
Let us write the static parameters as 
$r(\Lambda)$,  $K(\Lambda)$, $u(\Lambda)$,  $\gamma(\Lambda)$ and $C(\Lambda)$ to make
 their $\Lambda$ dependence, explicitly.
These variables have the following asymptotic behaviors \cite{onukin,onuki,glw}:
\begin{eqnarray}
r(\Lambda ) &&\sim\Lambda^{2-\eta}, \label{eq: rlambda}\\
K(\Lambda ) &&\sim\Lambda^{-\eta}, \\
u(\Lambda ) &&\sim\Lambda^{\epsilon-2\eta}, \\
\gamma(\Lambda ) &&\sim\Lambda^{(\epsilon +\alpha /\nu )/2-\eta}, \label{eq: glambda}\\
C(\Lambda ) &&\sim\Lambda^{-\alpha/\nu},  \label{eq: clambda}
\end{eqnarray}
where $\epsilon=4-d$ with $d$ being the space dimension, while $\alpha$, $\nu$ and $\eta$ 
are the usual static critical exponents.
Noting that $\eta$ is of order $\epsilon^2$ and very small, 
 we neglect $\eta$ and set $K_0=1$, hereafter.

\subsection{Streaming terms and  bare kinetic coefficients}

Here, we construct the streaming terms, $v_n$, $v_e$ and $\bfv_{J}$.
We can nicely determine the first two terms from the continuity equations, 
because $\delta n$ and $\delta e$ are the conserved densities.
From the continuity equations, we can write $v_n$ and $v_e$ as divergences of reversible currents,
which read
\begin{eqnarray}
\bfj_{n}=&&n\gamma \delta \bfv , \label{eq: jrn}\\
\bfj_{e}=&&(e+P)\gamma^2\delta\bfv ,\label{eq: jre}
\end{eqnarray}
with $\bfj_{n}$ and $\bfj_{e}$ being the reversible currents of the number  and energy density, respectively. 
Here, $\gamma$ is the Lorentz factor of the fluid-velocity fluctuation, $n=n_c+\delta n$ and $e=e_c+\delta e$.
As the reference frame, we chose the rest frame of the equilibrium state,
 and then the back ground fluid velocity vanishes.
Furthermore, We may set $\gamma \sim 1$,
because the fluid-velocity fluctuation is given by $\delta \bfv = h_c^{-1} \delta \bfJ$ that is not enhanced.
Therefore, we  write the streaming terms, $v_n$ and $v_e$, as
\begin{eqnarray}
v_n&&=-\nabla \cdot (n\delta \bfv), \label{eq: vn}\\
v_e&&=-\nabla \cdot ((e+P_c)\delta \bfv ) \label{eq: ve},
\end{eqnarray}
where we neglect the pressure fluctuation because
it is not enhanced near the critical point \cite{Minami:2009hn}.

Now, we note that the determination of $\bfv_{J}$ is not simple.
 Although the continuity equation tells us that 
 $\bfv_J$ is the divergence of the reversible-stress tensor,
 the determination of the reversible-stress tensor is not trivial. 
However,
we can determine it from the potential condition, 
which is a general condition for the streaming terms \cite{onuki}.  
The potential (or divergence) condition \cite{onuki, mazenko} reads
\begin{equation}
\int d\bfr \sum\limits_{j=n, e, J}  v_j(A) \frac{\delta(\beta H)}{\delta A_j }
 =\int d\bfr \sum\limits_{j=n, e, J}\frac{\partial v_j(A )}{\partial A_j }. \label{eq: potentialcon}
\end{equation}
We remark that this condition can be derived
 from Eq.(\ref{eq: vja}).
In a continuum system, the right-hand side of Eq. (\ref{eq: potentialcon})  vanishes in general \cite{onuki}. 

Thus,  the potential condition is reduced to 
\begin{equation}
\int d\bfr \sum\limits_{j=n, e, J}  v_j(A) \frac{\delta(\beta H)}{\delta A_j}=0, \label{eq: condition}
\end{equation}
where  $\bfv_J$ is only the unknown quantity 
because we have already determined $v_n$, $v_e$ and $H(\delta n, \delta e, \delta \bfJ)$.
Using Eqs. (\ref{eq: vf}), (\ref{eq: vn}), (\ref{eq: ve}) and (\ref{eq: condition}),
we obtain
\begin{eqnarray}
\int d\bfr \bl[ n\nabla \frac{\delta H}{\delta n}+(e+P_c)\nabla\frac{\delta H}{\delta e}+\bfv_J \br]
\cdot \beta \delta\bfv=0.
\end{eqnarray}
Because this condition should be satisfied for an arbitrary fluid-velocity fluctuation, 
we have 
\begin{equation}
\bfv_J =-n\nabla\frac{\delta H}{\delta n} - (e + P_c)\nabla \frac{\delta H}{\delta e}.
\end{equation}

Next, let us determine the kinetic coefficients from 
the relativistic hydrodynamic equation, Eqs. 
(\ref{eq: n})-(\ref{eq: d2}).
From Eqs. (\ref{eq: d1}), (\ref{eq: d2}), (\ref{eq: muf}) and (\ref{eq: vf}), 
we can read the kinetic coefficients $L_{j k} $ for small $\delta \bfv$ as
\begin{eqnarray}
L_{n n}&&=-\lambda_0\bl(\frac{n_c T_c}{h_c}\br)^2\nabla^2, \\
L_{J J}^{i j}&&=-T_c [\eta_0 \delta_{i j}\partial_i \partial_j
              +(\zeta_0 +(1-2/d)\eta_0)\partial_i\partial_j ],
\end{eqnarray}
and that the other coefficients are zero.
Here, $d$ is the space dimension and the transport coefficients, $\lambda_0, \eta_0$ and $\zeta_0$,
are {\em bare} ones.

Now,  we have determined all the terms, 
and then can write the nonlinear Langevin equation for the QCD critical point as 
\begin{eqnarray}
\frac{\partial \delta n}{\partial t}=&&-\nabla \cdot (n\delta \bfv)
                  -L_{n n}\frac{\delta (\beta H)}{\delta n}+\theta_n , \label{eq: nln} \\
\frac{\partial \delta e}{\partial t}=&&-\nabla \cdot ((e+P_c)\delta \bfv) ,\label{eq: nle} \\
\frac{\partial \delta \bfJ}{\partial t}=&&-n\nabla \frac{\delta H}{\delta n}
-(e+P_c)\nabla\frac{\delta H}{\delta e} -L_{J J} \cdot \frac{\delta(\beta H)}{\delta \bfJ}+ \bftheta_J ,\label{eq: nlj}
\end{eqnarray}
where $\theta_n$ and $\bftheta_J$ are the noise terms and satisfy the fluctuation-dissipation relations
\begin{eqnarray}
\la \theta_n (\bfr ,t) \theta_n (\bfr^{'} ,t^{'})\ra =&&-2\lambda_{0}
    \bl(\frac{n_c T_c}{h_c}\br)^2\nabla^2\delta (\bfr -\bfr^{'} ) \delta (t-t^{'}), \label{eq: fdrn} \\
\la \theta_J^i(\bfr,t) \theta_J^i(\bfr^{'},t^{'}) \ra =&&-2 T_c[\eta_0 \delta^{i j}\nabla^2 +\{\zeta_0+(1-2/d)\eta_0\}\partial^i\partial^j] \nonumber \\
   &&\times\delta(\bfr-\bfr^{'})\delta(t-t^{'}). \label{eq: fdrj}
\end{eqnarray}
Here, an important point is that the first and second terms in Eq. (\ref{eq: nlj}) denote
the nonlinear effects that are absent in the usual hydrodynamics.
These terms represent the time-reversible forces acting on the fluid, and
the force in the usual hydrodynamics is only the pressure gradient, $-\nabla P$.
Thus, the nonlinear part in those terms reflect the softening of the thermodynamic potential
and is not included in the hydrodynamics.

Let us now write the transport coefficients as $\lambda(\Lambda)$, $\eta(\Lambda)$ and $\zeta(\Lambda)$
to make their cutoff dependence in the critical region.
The critical behaviors of the transport coefficients
 are determined from their asymptotic behaviors near the relevant fixed point
as  $\Lambda$ is lowered.  

Now, we compare the Langevin equation, Eqs. (\ref{eq: nln}) - (\ref{eq: nlj}), 
with that for the liquid-gas critical point \cite{onukin}
\begin{eqnarray}
\frac{\partial \delta n}{\partial t}=&&-\nabla \cdot (n\delta v),\\
\frac{\partial \delta e}{\partial t}=&&-\nabla \cdot ((e+P_c)\delta v)
  +\lambda_0 T_c \nabla^2\frac{\delta H_{\rm lg}}{\delta e}+\theta_e ,\\
\frac{\partial \delta \bfJ_{\rho}}{\partial t}=&&-n\nabla \frac{\delta H_{\rm lg}}{\delta n}
  -(e+P_c)\nabla\frac{\delta H_{\rm lg}}{\delta e}-L_{J J} \cdot \frac{\delta(\beta H_{\rm lg})}{\delta \bfJ_\rho}+ \bftheta_J ,
\end{eqnarray}
where $\delta \bfJ_\rho \equiv \rho_c \delta \bfv$, 
$\rho$ and $H^{\rm lg}$ are  the non-relativistic momentum density, the mass density
 and the thermodynamic potential for liquid-gas critical point, respectively:  
\begin{equation}
H_{\rm lg}(\delta n,\delta e, \delta \bfJ_{\rho} )=H_{\rm Ising}(\psi, m) + \frac{1}{2 \rho_c}\delta \bfJ_{\rho}^2.  
\end{equation}
We see that the streaming terms have the same forms but
the dissipative ones are different between the relativistic and non-relativistic cases.
The difference also appears the relation between the momentum
and the fluid-velocity fluctuation.
Therefore, one may naturally  expect some novel characteristics  in the 
relativistic case that is absent in the non-relativistic case \cite{onukin}.

\section{The transport coefficients by dynamic RG }

Here, we  study the transport coefficients by the dynamic RG.
A detailed derivation of the RG equations is given in the Appendixes.

First, we rewrite Eqs. (\ref{eq: nln}) - (\ref{eq: nlj}) as the equation in terms of $(\psi , m)$,
to conform $( \delta n ,\delta e )$ to  $(\psi ,m)$.
Noting that we can set $\alpha_2=0$ in the mapping relations,
 Eqs. (\ref{eq: map1}) and (\ref{eq: map2}), without loss of generality \cite{onukin},
we have
\begin{eqnarray}
\frac{\partial \psi}{\partial t}
    =&&-C_{\psi} \nabla\cdot\delta \bfJ -\alpha_1^{-2}L_{n n}\frac{\delta(\beta H)}{\delta \psi}
         -h_c^{-1}\nabla\cdot (\psi \delta \bfJ) +\alpha_1^{-1}\theta_n ,\label{eq:psi} \\
\frac{\partial m}{\partial t} =&&-\beta_2^{-1}\nabla\cdot\delta \bfJ -h_c^{-1}\nabla\cdot(m\delta \bfJ) ,
  \label{eq:m} \\
\frac{\partial(\delta \bfJ)}{\partial t}
    =&&
      -C_J\nabla\frac{\delta H}{\delta \psi}-\beta_2^{-1}h_c\nabla\frac{\delta H}{\delta m}
     \nonumber \\
      &&-\psi \nabla \frac{\delta H}{\delta \psi}-m\nabla\frac{\delta H}{\delta m} 
     -(T_c h_c)^{-1}L_{J J}\cdot \delta \bfJ +\bftheta_J ,
\label{eq:J}
\end{eqnarray}
with
$C_{\psi} \equiv \alpha_1^{-1}(n_c h_c^{-1}-\beta_1\beta_2^{-1})$ and
$C_{J}\equiv \alpha_1^{-1}(n_c-\beta_1 h_c)$. 
Here, we note that we could rewrite the potential, Eq.(\ref{eq: glw}), 
as that in terms of $(\delta n, \delta e)$ to conform the variables;
the choice is a matter of preference.

In the dynamic RG transformation, 
we average over the short wavelength components in the shell, 
$\Lambda-\delta\Lambda < k < \Lambda$, for the Langevin equation.
For this task, we must perturbatively solve the equation about them,
by rewriting it as a self-consistent equation \cite{mazenko}.    
Although an explicit derivation of the self-consistent equation 
for the QCD critical point is first made in this thesis,
 we leave the details of the derivation to Appendix \ref{sec: derivation},
because the general procedure of the derivation is
standard and given in the textbook \cite{mazenko}.
Here, we provide only a few basic equations of the dynamic RG for the QCD critical point.

Now, as is shown in Appendix \ref{sec: derivation},
 Eqs. (\ref{eq:psi})-(\ref{eq:J}) can be reduced to the following form; 
\begin{equation}
  \begin{pmatrix}
  \tilde{\psi}(\bfk ,\omega) \\
  \tilde{m}(\bfk ,\omega) \\
  \delta \tilde{J}_{\parallel}(\bfk,\omega)
  \end{pmatrix}
=
 \begin{pmatrix}
  \tilde{\psi}^0(\bfk ,\omega) \\
  0 \\
  \delta \tilde{J}_{\parallel}^0(\bfk,\omega)
  \end{pmatrix}
+G^0(\bfk,\omega)\bfV(\bfk,\omega),
\label{eq: self1}
\end{equation}
and
\begin{equation}
\delta \tilde{\bfJ}_{\perp}(\bfk,\omega)
=\delta \tilde{\bfJ}_{\perp}^0(\bfk,\omega)+G^0_{\perp}\bfV_{\perp \psi\psi}(\bfk,\omega),
\label{eq: self2}
\end{equation}
where $\delta\tilde{J}_{\parallel }(\bfk) \equiv \hat{\bfk} \cdot \delta\tilde{J}(\bfk)$ and 
$\delta\tilde{J}_{\perp}(\bfk) \equiv \delta \tilde{J}(\bfk) - \delta \tilde{J}_{\parallel}(\bfk)$ are 
the longitudinal and  transverse components of the momentum.  
Here, $G^0$ and $G^0_{\perp}$ are the bare propagators, which are given by Eqs. (\ref{eq: g0}) and 
 (\ref{eq: gthermal}) - (\ref{eq: gviscous}),
whereas $\bfV$ and $\bfV_{\perp \psi\psi}$ the nonlinear couplings, coming from the streaming terms
 and given by Eqs. (\ref{eq: v1}) - (\ref{eq: v3}) and (\ref{eq: v4}). 
Also, $\tilde{\psi}^0$, $\delta \tilde{J}_{\parallel}^0$ and $\delta \tilde{\bfJ}_{\perp}^0$
are the bare variables, which are the solutions without the nonlinear terms.  
Iterating the self-consistent equations (\ref{eq: self1}) and (\ref{eq: self2}), 
we can obtain a perturbative expansion of the nonlinear couplings
and have a coarse-grained Langevin equation.

Now, we note that the variables, $\psi $,$\tilde{J}_{\perp}$ and $\tilde{J}_{\parallel} $,
are respectively correspond to the thermal,  viscous and  sound modes
\footnote{Although $\tilde{m}$ would be a linear combination of the thermal and  sound modes,
we need not to consider $\tilde{m}$ for a following analysis. } 
(see the propagators  (\ref{eq: gthermal}) - (\ref{eq: gviscous}).).
Thus, the first and third rows of Eq. (\ref{eq: self1}) 
respectively denote the equations of  motion for the thermal and sound modes,
while Eq. (\ref{eq: self2})  for the viscous mode. 
We stress that the sound mode is neglected in the model H,
although it is essential for the renormalization of the bulk viscosity.    

Here, we  make the coarse graining 
to the second order in the nonlinear couplings,  $\bfV$ and $\bfV_{\perp \psi\psi}$
(see Fig. \ref{fig: cgpeom} for an example.).
Inspecting the coarse-grained equation for $\tilde{\psi}$ 
(see Eq. (\ref{eq: cgpeom}) for the detail),
we have the RG equation for the thermal conductivity:
\begin{eqnarray}    
-\Lambda\frac{\partial \lambda(\Lambda)}{\partial \Lambda}=&&\frac{3}{4}f(\Lambda)\lambda(\Lambda),
\label{eq: rgl} 
\end{eqnarray}
$f(\Lambda)\equiv T_c K_4/(D_{\psi}\eta(\Lambda)\lambda(\Lambda)\Lambda^{e}) $,
$K_4$ is the surface area of a unit sphere in 4 dimensions divided by $(2\pi)^4$,
$D_{\psi}\equiv (n_c T_c /\alpha_1 h_c)^2$.
Here, we have introduced $f(\Lambda)$ for convenience sake.
Similarly, from the coarse-grained equations for $\tilde{J}_{\perp} $ and $\tilde{J}_{\parallel} $,
we obtain the RG equations for the shear and bulk viscosities
\begin{eqnarray}  
-\Lambda\frac{\partial \eta(\Lambda)}{\partial \Lambda}=&&\frac{1}{24}f(\Lambda)\eta(\Lambda),
\label{eq: rge} \\
-\Lambda\frac{\partial \zeta(\Lambda)}{\partial \Lambda}=&&
A\gamma^2(\Lambda)\lambda^{-1}(\Lambda)\Lambda^{-\epsilon-4},
\label{eq: rgz}
\end{eqnarray}
where  $\gamma(\Lambda)$ is a static parameter in the thermodynamic potential 
(see Eqs. (\ref{eq: glw}) and (\ref{eq: glambda})), and $A \equiv  h_c^2 K_4/(\beta_2^2 D_{\psi})$.
Furthermore, differentiating $f(\Lambda)$ about $\Lambda$,
we also have the RG equation for it:
\begin{equation}
-\Lambda\frac{\partial f(\Lambda)}{\partial \Lambda}=f(\Lambda )\bl( \epsilon-\frac{19}{24}f(\Lambda ) \br).
\label{eq: rgf} 
\end{equation}

Now, we note that Eqs. (\ref{eq: rgl}), (\ref{eq: rge}) and (\ref{eq: rgf}) are identical to those for  the
liquid-gas critical point except for unimportant constants in $f(\Lambda)$\cite{onuki, onukin}.
Equation (\ref{eq: rgz}) is also equivalent 
to the RG equation of the bulk viscosity for the liquid-gas critical point in the limit $\omega \rightarrow 0$ 
\cite{onuki, onukin}. 
Therefore, arguments about the RG equations and results from those are the same 
as for the liquid-gas critical point.
Then, we provide only essential  arguments and results in the following part,
and leave the detail to \cite{onuki, kg, siggia, onukin}.

Now, we identify the relevant-fixed point as the following \cite{onuki, onukin}.
Because, at a fixed point, parameters are invariant about the RG transformation,
we set the left-hand side of Eq. (\ref{eq: rgf}) as $0$.
Then, as a fixed-point value of $f(\Lambda)$ which is denoted by $f^{*}$,
 we have $f^*=0$ and $f^{*}=(24/19)\epsilon$.
Therefore, we have the two fixed points, and the relevant one is specified by $f^{*}=(24/19)\epsilon$.
Although the relevant point seems to be absent in Eqs. (\ref{eq: rgl}), (\ref{eq: rge}) and (\ref{eq: rgz}),
 the reason is due to the simplified RG transformation as mentioned in the earlier section,
and this is just a apparent problem \cite{kg, siggia}.   

Substituting $f^{*}=(24/19)\epsilon$ into Eqs. (\ref{eq: rgl}), (\ref{eq: rge}) and (\ref{eq: rgz}),
we have the asymptotic behaviors near the relevant-fixed point:
\begin{eqnarray}
\lambda(\Lambda) \sim && \Lambda^{-\frac{18}{19}\epsilon}, \\
\eta(\Lambda) \sim && \Lambda^{-\frac{1}{19}\epsilon}, \\
\zeta(\Lambda) \sim && \Lambda^{-(4-\frac{18}{19}\epsilon-\frac{\alpha}{\nu})} \label{eq: zlambda}.
\end{eqnarray} 
Here, in the derivation of Eq. (\ref{eq: zlambda}), we have used the asymptotic behavior of $\gamma(\Lambda)$,
 Eq. (\ref{eq: glambda}).
Decreasing the cutoff to the region $\Lambda \ll \xi^{-1}$, 
we can replace $\Lambda$ with $ \xi^{-1}$ in the asymptotic behaviors \cite{onuki, siggia}:
\begin{eqnarray}
\lambda_{\rm R}\sim && \xi^{\frac{18}{19}\epsilon}, \label{eq: lr}\\
\eta_{\rm R} \sim && \xi^{\frac{1}{19}\epsilon}, \label{eq: er}\\
\zeta_{\rm R} \sim && \xi^{4-\frac{18}{19}\epsilon-\frac{\alpha}{\nu}}. \label{eq: zr}
\end{eqnarray} 
In three dimensions, we find
\begin{eqnarray}
\lambda_{\rm R}\sim && \xi^{0.95}, \\
\eta_{\rm R} \sim && \xi^{0.053}, \\
\zeta_{\rm R} \sim && \xi^{2.8}.
\end{eqnarray} 

We can also read the dynamic critical exponents from Eqs. (\ref{eq: lr})-(\ref{eq: zr}).
A dynamic critical exponent, denoted by $z$, generally parametrizes 
the decay rate $\Gamma(k)$ at the wavenumber $k=\xi^{-1}$ as 
$\Gamma (\xi^{-1}) \sim \xi^{-z}$.
As shown in Appendix \ref{sec: derivation},
the decay rates for the three modes 
at $k$ are given by
\begin{eqnarray}
\Gamma_{\rm thermal}(k) =&& \lambda_{\rm R} k^2 (r_{\rm R} +k^2) D_{\psi}, \\
\Gamma_{\rm viscous}(k) =&& \eta_{\rm R} k^2 h_c^{-1}, \\
\Gamma_{\rm sound}(k) =&&  (\zeta_{\rm R} +2(1-1/d)\eta_{\rm R}) k^2 h_c^{-1}.
\end{eqnarray}  
Thus, we find the dynamic critical exponents as
\begin{eqnarray}
z_{\rm thermal} =&&4-\frac{18}{19}\epsilon, \\
z_{\rm viscous} =&&2-\frac{1}{19}\epsilon, \\
z_{\rm sound} =&&- \bl( 2-\frac{18}{19}\epsilon-\frac{\alpha}{\nu} \br) .
\end{eqnarray}
In three dimensions, those are
\begin{eqnarray}
z_{\rm thermal} \sim &&3, \\
z_{\rm viscous} \sim &&2, \\
z_{\rm sound} \sim &&-0.8.
\end{eqnarray}
We see that the thermal and  viscous modes exhibit critical slowing down, while
the sound mode critical speeding up.

Why do not the relativistic effects appear in the RG equations?
The reason is that
the nonlinear terms in the dissipative terms 
generally renormalize only static parameters, 
up to order $\epsilon^2$ \cite{siggia, mazenko}.
Furthermore,
the difference in the relation between the momentum
and the fluid velocity is only unimportant constants, 
i.e., the enthalpy density $h$ and the mass density $\rho$.
Then, the RG equations are essentially the same as for the non-relativistic case.

\chapter{Summary and Concluding remarks}

We studied the critical dynamics near the QCD critical point 
by the linearized relativistic hydrodynamics and the dynamic renormalization group (RG).

First, we studied the linear dynamics of the baryon number and the energy-momentum 
in the hydrodynamic regime, $k \ll \xi^{-1} $, by the relativistic hydrodynamics.
We showed that the actual slow modes are  three:
the thermal, viscous and sound modes.
Furthermore, we found that, near the critical point, 
the thermal mode is enhanced and the most relevant.
In contrast, the sound is suppressed and negligible for a minimal critical dynamics.

By this study, We also found that 
the Landau equation, which is believed to be an acausal hydrodynamic equation,
has no problem to describe the slow dynamics.
We also showed that the Israel-Stewart equation, which has relaxation time and is causal,
gives the same slow dynamics as the Landau equation gives.
This result implies that the causality problem occurs only the short-wavelength region.
We note that the short-wavelength region would be out of applicable scope for the hydrodynamics
and actually ruled out by the cutoff. 
We also stress that
the relaxation time is important only for a rapid motion.

Next, we  studied the nonlinear dynamics in the critical regime, $\xi^{-1} \ll k$,
by the dynamic RG.
For this purpose, we constructed
 the nonlinear Langevin equation  near  the critical point for the first time. 
Our construction is 
based on the generalized Langevin theory, by 
Mori \cite{mori,mori:fujisaka}, and the relativistic hydrodynamics;
instead of a naive construction method \cite{mazenko},
we determined  the
streaming terms by  the relativistic hydrodynamics
and the potential condition, which gives a constraint to these terms.  
The resulting equation is given by Eqs. (\ref{eq: nln})-(\ref{eq: nlj}). 
Although there are some attempts to 
make a one-to-one mapping between the QCD critical point and the Ising critical point \cite{moore, mapping},
we showed that it  is not necessary to specify such the mapping for the 
critical exponents, as for the liquid-gas critical point \cite{onuki}.    

We showed that 
the bulk viscosity and the thermal conductivity strongly diverge  at the QCD critical point. 
Also, we found that the thermal and viscous modes 
exhibit critical slowing down with the dynamic critical
exponents $z_{\rm thermal}\sim 3$  and $z_{\rm viscous}\sim 2$, respectively.  
In contrast,  the sound mode critical-speeding up
with the negative exponent $z_{\rm sound}\sim -0.8$.   
We stress that the earlier studies \cite{karsch, sasaki} 
treat the bare transport coefficients and does not include
the macroscopic nonlinear interaction.

We note that the bulk viscosity and the thermal conductivity
are usually neglected in  heavy ion physics, however
they become much more important than the shear viscosity near the QCD critical point.
Furthermore,  the description for the created matter as a perfect fluid is
not valid near the QCD critical point by the strong divergence of the bulk viscosity.

As the argument about the dynamic universality class \cite{moore, son},
we  showed, from an explicit calculation, 
that the QCD critical point has the same critical behaviors as the liquid-gas critical point has. 
The argument assumes the insignificance of the relativity for the critical dynamics by
 the slowness of the diffusion processes.
However, 
we  showed that the genuine reason for the insignificance
originates from the small fluctuation of the momentum density;
the critical dynamics is essentially governed by the streaming terms,
which are modified by  the relativistic effect
through only a Lorentz factor of the fluid velocity fluctuation.
However, the fluid-velocity fluctuation, which is proportional to the momentum, 
is not enhanced near the critical point.
Thus, the relativistic effects do not affect the critical 
dynamics near the critical point.
We stress that the sound mode exhibit critical speeding up,
and then the sound diffusion is fast near the critical point. 
Therefore, the basis of the conjecture would be true for the thermal and  viscous modes,
but not for the sound mode.
We also note that the model H \cite{kawasaki1},
which is the {\em minimal}-dynamic model for the dynamics near the liquid-gas critical point,
 can not describe the critical behavior of the bulk viscosity 
because it does not contain the sound mode.

We note that our Langevin equation must satisfy usual fluctuation-dissipation relations, 
Eqs. (\ref{eq: fdrn}) and (\ref{eq: fdrj}), for the consistency with the linearized Landau equation
\footnote{If our nonlinear Langevin equation is linearized, 
the linearized equation must give the same result as the Landau equation gives.},
although a relativistic Brownian motion seems not  to satisfy the usual relations \cite{Akamatsu:2008ge}.
Moreover, our Langevin equation seems to violate the causality, 
because the dissipative terms are determined from the Landau equation.
However,  the Israel-Stewart equation, in which the causality problem is formally resolved,
 gives the same result as the Landau equation gives in long-wavelength region,
as shown in Sec.\ref{sec: is}.  
Therefore, our determination from the Landau equation must suffice.

Also, we note a frame dependence of the results in Chap. \ref{sec:drgqcd}.
As a  hydrodynamic equation, we used only the equation in the energy frame.
Does the results change if an equation in the particle frame is used?
Although the frame dependence can appear in only dissipative terms,
 the critical dynamics is essentially determined by the streaming terms.
Therefore, the results would not change for the particle frame, if an equation in the frame is correct. 
However, in practice, the Eckart equation has a pathological behavior\cite{hiscock}.
Namely, fluctuations do not relax, and therefore we cannot use the Eckart equation.

Furthermore, we note Lorentz covariance.
Our Langevin equation is not Lorentz covariant,
but it would not be a problem.  The reason is the following.
Here, we consider the fluctuations in the background medium. 
In such situation, the Lorentz transformation boosts the fluctuations but not the medium.
Then, after the Lorentz boost, we have the boosted fluctuation and the medium that still rests.
Namely, the boosted system differs from that before the boost. 
The covariance means that, if we see the same system from different reference frames,
we have the same physics.
Thus, our Langevin equation would be no problem.
Actually, an equation of the Brownian motion is not Galilei covariant by the same reason,
but it is no problem.

Recently, some authors suggest the existence of other critical points in higher density region
of the QCD phase diagram where the color superconductivity is taken into account
\cite{Kitazawa:2002bc,Hatsuda:2006ps}.
It would be interesting to study the critical dynamics near  such a new QCD critical point using the dynamic RG theory,
as an extension of the present work.
For this purpose, however, we must firstly specify the soft modes and 
construct the nonlinear Langevin equation.
If the soft modes are different from the conserved densities, which is the case when
the  diquark fluctuations are relevant \cite{Kitazawa:2002bc,Kitazawa:2001ft,Alford:2000ze},
the construction based on the relativistic hydrodynamics done in the present work does not work,
and we must directly recourse to  Eq. (\ref{eq: stream}) to identify  the streaming terms.

\section*{ACKNOWLEDGMENTS}
I am grateful to  Yoshimasa Hidaka.

\appendix
\chapter{Generalized nonlinear Langevin equation}
\label{sec:a1}

Here, we give the detailed
derivation of the nonlinear Langevin equation (\ref{eq: nllangevin}).
As mentioned in \ref{sec:nle},  we can obtain the nonlinear Langevin equation about $A_j$
from the linear Langevin equation about $g(A,a)$ \cite{onuki, mazenko}.

Thus, let us first derive the linear equation about $g(A,a)$.
We note that the operator identity Eq.(\ref{eq: oi}) is valid even for ${\cal P}_g$.
Multiplying Eq.(\ref{eq: oi}) by $g(A,a)$, 
we have the linear Langevin equation for $g(A(t), a) \equiv \exp[i{\cal L} t] g(A, a)$:
\begin{equation}
\frac{\partial}{\partial t}g(A(t), a)=
  \int da^{'} i\Omega_{a a^{'}}g(A(t),a^{'})
  -\int^t_0 dt\int da^{'}\Psi_{a a^{'}}(t^{'})g(A(t-t^{'}),a^{'})+F_a(t). \label{eq: gnl}
\end{equation}
Here, 
\begin{eqnarray}
i\Omega_{a a^{'}}&&=\la (iL\delta(A-a) ) \delta(A-a^{'})\ra/P_{eq}(a^{'})  , \\
                 &&=-\sum\limits_{j}\frac{\partial}{\partial a_j}[v_j(a)\delta(a-a^{'})],
\end{eqnarray}
where $v_j (a)$ is 
\begin{equation}
v_j(a)=\la \dot{A}_j \delta(A-a)\ra /P_{eq}(a)=\la \dot{A}_j; a\ra .
\end{equation}
The noise $F_a (t)$ is 
\begin{eqnarray}
F_a(t)&&=\exp[{\cal Q}_g iLt]{\cal Q}_g iL\delta(A-a)  \\
      &&=-\sum\limits_k \frac{\partial}{\partial a_k}[\exp{[{\cal Q}_g iLt]}({\cal Q}_g\dot{A}_k)\delta(A-a)] \\
      &&=-\sum\limits_k \frac{\partial}{\partial a_k}[U(t)\theta_k(0) \delta (A-a)],
 \label{eq:nf}
\end{eqnarray}
where we  introduced
\begin{eqnarray}
U(t)&&=\exp{[{\cal Q}_g iLt ] ,} \\
\theta_j(t) &&=U(t){\cal Q}_g iL A_j(0) .
\end{eqnarray}
With the noise,  the memory function $\Psi_{a a^{'}}(t)$ is written as 
\begin{equation}
\Psi_{a a^{'}}(t)=\la F_a(t)F_{a^{'}}(0) \ra / P_{eq}(a^{'}).
\end{equation}
If we multiply Eq.(\ref{eq: gnl}) by $a_j$ and integrate over $a$, 
we can have an exact relation corresponding to Eq.(\ref{eq: gllangevin}).

However, let us now make a Markov approximation
because the exact expression is inconvenient for our purpose.
Again, if the time scale of $A_j$ is much larger than that of $\theta_j$,
we can make  the approximations: 
\begin{equation}
F_a(t) \sim - \sum\limits_j \frac{\partial}{\partial a_j}[\theta_j (t) \delta (A-a)], \label{eq: kinji}
\end{equation} 
and 
\begin{equation}
\int^t_0 dt\int da^{'}\Psi_{a a^{'}}(t^{'})g(A(t-t^{'}),a^{'}) 
\sim \int da \bl[ \int^{\infty}_0dt^{'}\Psi_{a a^{'}}(t^{'}) \br] g(A(t),a^{'}).
\end{equation}
Introducing a bare transport coefficients as,
\begin{equation}
L_{i k}(a) \equiv \int^{\infty}_0dt\la \theta_i (t) \theta_k (0);a\ra .
\end{equation}
we can rewrite the memory function as
\begin{eqnarray}
\int^{\infty}_0 dt\Psi_{a a^{'}}(t)&&=\frac{1}{P_{\rm eq}(a^{'})}\int^{\infty}_0 dt\la F_a(t) F_{a^{'}}(0)\ra ,\\
  && = \frac{1}{P_{eq}(a^{'})}\sum\limits_{i k}\frac{\partial}{\partial a_i}\frac{\partial}{\partial a^{'}_k}
[L_{i k}(a) P_{eq}(a)\delta(a-a^{'})].
\end{eqnarray}
After these approximations, we can obtain the nonlinear Langevin equation (\ref{eq: nllangevin})
from Eq.(\ref{eq: gnl}).

\chapter{Detailed derivation in the chapter \ref{sec:sdaqcp} }

\label{sec:dds}

Here, we gives the detailed derivation of Eq.(\ref{eq:landau}).

Let us first calculate the matrix elements, $(A^{-1})_{1 1}$ and $(A^{-1})_{1 3}$,
in Eq.(\ref{eq:tilden}).
They are given by  the simple formula
$(A^{-1})_{1 1}=\frac{1}{\det A}(A_{2 2}A_{3 3}-A_{2 3}A_{3 2})$
and $(A^{-1})_{1 3}=\frac{1}{\det A}(A_{1 2}A_{2 3}-A_{1 3}A_{2 2})$.
Here, $\det A$ reads 
\begin{eqnarray}
\det A =\frac{n_c c_v}{h_c T_c}
         \bl[ z^3 
          &+&z^2 k^2 \bl( \frac{c_p}{c_v} D_{\rm t} +\nu_l +\lambda\frac{T_c c_s^2}{h_c }
         -2 D_{\rm t} c_s^2 \alpha_P T_c \br)                  \nonumber  \\
          &+&z k^2 c_s^2
          +k^4 c_s^2 D_{\rm t}
          +...  \br],
\end{eqnarray}
where  '$...$' denotes the higher order terms in $k$.
We are interested in the low-wavenumber region.
Then,
$\det A$ can be nicely factorized to second order in $k$,
\begin{equation}
\det A \sim \frac{n_c c_v}{h_c T_c}
(z+D_{\rm t} k^{2})(z+D_{\rm s} k^{2}+i c_{s}k)(z+D_{\rm s} k^{2}-i c_{s}k),
\end{equation}
where 
\begin{eqnarray}
D_{\rm s} &=& \frac{1}{2}\bl[ D_{\rm t} \bl( \frac{c_p}{c_v} -1\br)+\nu_{l}
+c_s^2 T_c \bl( \lambda / h_c -2 D_{\rm t} \alpha_P \br) \br].
\end{eqnarray}

Then, we can write the Fourier-Laplace coefficient of the density fluctuation
to second order in $k$, 
\begin{equation*}
\frac{\delta n(\bfk ,z)}{\delta n(\bfk ,0)} \sim 
\frac{(z+D_{\rm s} k^{2}+i c_{s}k)(z+D_{\rm s} k^{2}-i c_{s}k)
+z k^{2}\frac{\lambda c_s^2 c_v}{h_c c_p}
-k^2\frac{c_s^2 c_v}{c_p}}
{(z+D_{\rm t} k^{2})(z+D_{\rm s} k^{2}+i c_{s}k)(z+D_{\rm s} k^{2}-i c_{s}k) }.
\end{equation*}

Performing the inverse Laplace transformation 
\begin{equation}
\delta n(\bfk ,t)/\delta n(\bfk ,0)=\frac{1}{2\pi i} \int ^{\delta+i\infty }_{\delta-i\infty}dz
\e ^{zt}\delta n(\bfk ,z)/\delta n(\bfk ,0),
\end{equation}
we obtain the dynamical density fluctuation at $t$
\begin{equation}
\delta n(\bfk ,t)/\delta n(\bfk ,0) \sim  
\bl( 1-\frac{c_v}{c_p} \br)
\e^{- D_{\rm t} k^{2}t} + \frac{c_v}{c_p}
\cos (c_s k t) \e^{-D_{\rm s} k^{2} t}. 
\label{eq:tdensity}
\end{equation}
Here, we have retained only the terms in the amplitudes to zeroth order in $k$.
Because Eq.(\ref{eq:tdensity}) is the density fluctuation in a stationary process, 
we can replace the time $t$ by $\vert  t \vert $.
Therefore the Fourier transformation of Eq.(\ref{eq:tdensity}) is 
\begin{eqnarray}
\delta n(\bfk ,\omega )/\delta n(\bfk ,0)&&=
\bl( 1-\frac{c_v}{c_p} \br)
\frac{2 D_{\rm t} k^{2}}{\omega^{2}+D_{\rm t}^{2}k^{4}}
 \nonumber \\
+&&\frac{c_v}{c_p}
\bl[ \frac{D_{\rm s} k^{2}}{(\omega -c_{s}k)^{2}+D_{\rm s}^{2}k^{4}}
+\frac{D_{\rm s} k^{2}}{(\omega +c_{s}k)^{2}+D_{\rm s}^{2}k^{4}} \br].
\end{eqnarray}
Thus, we finally obtain the spectral function of the density fluctuation  
\begin{eqnarray}
S_{n n}(\bfk ,\omega ) &=& \la \delta n(\bfk ,\omega ) \delta n(\bfk ,t=0)\ra \nonumber \\
   &=& \la (\delta n(\bfk ,t=0))^2\ra \bl[\; \bl( 1-\frac{c_v}{c_p} \br)
   \frac{2 D_{\rm t} k^{2}}{\omega^{2}+D_{\rm t}^{2}k^{4}}
   \nonumber \\
   &+&\frac{c_v}{c_p}
   \bl( \frac{D_{\rm s} k^{2}}{(\omega -c_{s}k)^{2}+D_{\rm s}^{2}k^{4}}
   +\frac{D_{\rm s} k^{2}}{(\omega +c_{s}k)^{2}+D_{\rm s}^{2}k^{4}} \br) \;\br].
\end{eqnarray}

\chapter{Rewriting the nonlinear Langevin equation as a self-consistent equation}

\label{sec: derivation}
Here, we rewrite the Langevin equation, Eqs. (\ref{eq:psi})-(\ref{eq:J}) as a self-consistent equation.
First, we make a Fourier transformation as the following
\begin{eqnarray}
\tilde{\psi}(\bfk,\omega) =\int dt d^d r e^{i\omega t-i\bfk\cdot\bfr} \psi(\bfr,t). \label{eq: foo}
\end{eqnarray}
Then, we have
\begin{eqnarray}
-i\omega\tilde{\psi}(\bfk,\omega)
 =&&- C_{\psi} i\bfk \cdot\delta\tilde{\bfJ}
    -\alpha_1^{-2}\tilde{L_{n n}}\frac{\delta(\beta \tilde{H})}{\delta \psi} \nonumber \\
    &&-h_c^{-1} i\bfk \cdot \int_{q \Omega} (\tilde \psi(q) \delta \tilde{\bfJ}(k-q)) +\alpha_1^{-1}\tilde{\theta_n}, 
  \label{eq: fpsi}\\
-i\omega \tilde{m}(\bfk,\omega)
  =&&-\beta_2^{-1}i\bfk\cdot\delta \tilde{\bfJ}
    -h_c^{-1}i\bfk\cdot\int_{q\Omega}(\tilde{m}(q)\delta \tilde{\bfJ}(k-q)), \\
-i\omega\delta\tilde{\bfJ}(\bfk,\omega ) 
 =&&
 -C_J i\bfk\frac{\delta \tilde{H}}{\delta \psi}-\beta_2^{-1}h_c i\bfk\frac{\delta \tilde{H}}{\delta m}
 \nonumber \\
    &&-i\int_{q\Omega}\bfq  \bl[ \frac{\delta \tilde{H}}{\delta \psi}(q)\tilde{\psi}(k-q)
                                +\frac{\delta \tilde{H}}{\delta m}(q)\tilde{m}(k-q)\br] \nonumber \\ 
    &&-(T_c h_c)^{-1}\tilde{L}_{J J}\cdot \delta \tilde{\bfJ} +\tilde{\bftheta_J}. \label{eq: fj}
\end{eqnarray}
Note that the quantities with tilde in Eq. (\ref{eq: fpsi})-(\ref{eq: fj}) are Fourier transformed,
like Eq. (\ref{eq: foo}), and we have abbreviated the nonlinear terms such as 
\begin{equation}
\int_{q\Omega}\tilde{\psi}(q)\delta\tilde{\bfJ}(k-q)
\equiv \int\frac{d \Omega}{2\pi}\frac{d^d q}{(2\pi)^d}\tilde{\psi}(\bfq,\Omega)\delta\tilde{\bfJ}(\bfk-\bfq,\omega-\Omega).
\end{equation}
We now decompose Eq. (\ref{eq: fj}) into the longitudinal and the transverse components:
\begin{eqnarray}
-i\omega\delta\tilde{\bfJ}_{\parallel}
   =&&-i C_J k \frac{\delta \tilde{H}}{\delta \psi}-i\beta_2^{-1}h_c k\frac{\delta \tilde{H}}{\delta m} \nonumber \\
   &&-i\int_{q\Omega}(\hat{\bfk}\cdot\bfq)  \bl[\frac{\delta \tilde{H}}{\delta \psi}(q)\tilde{\psi}(k-q)
                                +\frac{\delta \tilde{H}}{\delta m}(q)\tilde{m}(k-q)\br]  \nonumber \\
   &&-(T_c H_c)^{-1}\hat{\bfk}\cdot\tilde{L}_{J J}(\bfk )\cdot\delta\tilde{\bfJ}+\tilde\theta_{\parallel} \label{eq: parallel}, \\
-i\omega\delta \tilde{\bfJ}_{\perp}
   &&=-i\int_{q\Omega}{\cal P}_{\perp}(\bfk) \cdot\bfq \bl[\frac{\delta \tilde{H}}{\delta \psi}(q)\tilde{\psi}(k-q)
                                +\frac{\delta \tilde{H}}{\delta m}(q)\tilde{m}(k-q)\br]\nonumber \\ 
   &&   -(T_c h_c)^{-1}{\cal P}_{\perp}(\bfk) \cdot \tilde{L}_{J J}(\bfk) \cdot \delta \tilde{\bfJ}
    +\tilde{\bftheta}_{\perp} \label{eq: perp},
\end{eqnarray}
where we introduced a projection operator  as
\begin{equation}
({\cal P}_{\perp}(\bfk))_{i j}=\delta_{i j }-k_i k_j /k^2,
\end{equation}
and
\begin{eqnarray}
\delta \tilde{J}_{\parallel}(\bfk) &&=\hat{\bfk} \cdot \delta \tilde{\bfJ}(\bfk), \\
\delta \tilde{\bfJ}_{\perp}(\bfk) &&={\cal P}_{\perp}(\bfk) \cdot \delta \tilde{\bfJ} (\bfk), \\
\tilde{\theta}_{\parallel}(\bfk) &&=\hat{\bfk}\cdot\tilde{\bftheta}(\bfk),\\
\tilde{\bftheta}_{\perp}(\bfk) &&= {\cal P}_{\perp}(\bfk) \cdot \tilde{\bftheta}(\bfk).
\end{eqnarray}
Because the streaming terms in Eqs. (\ref{eq: parallel}) and (\ref{eq: perp}) are too complicated
for our purpose,
let us retain only the terms that yield dominant contributions for the transport coefficients.
We note that only such terms suffice for obtaining the critical exponents.
From the relations\cite{onuki} 
\begin{eqnarray}
\int d^3 r \la\psi(\bfr)\psi(0) \ra &&\sim \xi^2 ,\\
\int d^3 r \la m(\bfr) m(0) \ra && \sim \xi^{0.2} ,
\end{eqnarray}
we expect $\psi $ yields stronger singularity than $m$.
Therefore, we only retain the term that are of the second order in $\psi$.
Namely, we reduce the streaming terms as
\begin{eqnarray}
&&i C_J k \frac{\delta \tilde{H}}{\delta \psi}+i\beta_2^{-1}h_c k\frac{\delta \tilde{H}}{\delta m}
\nonumber \\
     && \hspace{0.5cm} +i\int_{q\Omega}(\hat{    \bfk}\cdot\bfq) 
   \bl[\frac{\delta \tilde{H}}{\delta \psi}(q)\tilde{\psi}(k-q)
    +\frac{\delta \tilde{H}}{\delta m}(q)\tilde{m}(k-q)\br]
     \nonumber \\
     &&\sim T_c \bl[ i C_J k\chi_0^{-1}(k)\tilde{\psi}+i\beta_2^{-1}h_c k C_0^{-1}\tilde{m}
     \nonumber \\
     &&\hspace{0.5cm} +i\beta_2^{-1}h_c k\gamma_0 \int_{q\Omega}\tilde{\psi}(q)\tilde{\psi}(k-q)\br] , \\
&&i\int_{q\Omega}{\cal P}_{\perp}(\bfk) \cdot\bfq  \bl[\frac{\delta \tilde{H}}{\delta \psi}(q)\tilde{\psi}(k-q)
     +\frac{\delta \tilde{H}}{\delta m}(q)\tilde{m}(k-q)\br] \nonumber \\
     &&\sim  i T_c{\cal P}_{\perp}(\bfk)\cdot\int_{q\Omega} \bfq \chi_{0}^{-1}(\bfq) \tilde{\psi}(q)\tilde{\psi}(k-q), 
\end{eqnarray}
where $\chi_0^{-1}(\bfk)=r_0+ k^2 $.
Notice that we set $K_0=1$, as mentioned in the text.

Next, we consider the dissipative terms.
The important point is that the nonlinear terms in dissipative terms generally renormalize 
 only static parameters in a thermodynamic potential, up to second order in $\epsilon$
\cite{siggia, mazenko}.
Therefore,
we can take into account  nonlinear terms in the dissipative terms
with the results of static RG, Eq. (\ref{eq: rlambda})-(\ref{eq: clambda}),
and effectively neglect it in the Langevin equation.
Then, we reduce the $\tilde{L}_{n n}\delta(\beta \tilde{H})/\delta \psi$ as
\begin{equation}
\tilde{L}_{n n}(\bfk )\frac{\delta \tilde{H}}{\delta \psi}(\bfk ,\omega)
\sim \lambda_0 k^2 \chi_0^{-1} \bl(\frac{n_c T_c}{h_c}\br)^2 \tilde{\psi}(\bfk,\omega).
\end{equation}
In contrast, the dissipative terms of $\delta \bfJ$ are originally linear and then directly read
\begin{eqnarray}
\hat{\bfk} \cdot \tilde{L}_{J J}(\bfk) \cdot \delta \bfJ(\bfk,\omega) 
  &&= T_c[\zeta_0+2(1-1/d)\eta_0 ]  k^2\delta\tilde{J}_{\parallel}(\bfk,\omega) ,\\
{\cal P}_{\perp} (\bfk)  \cdot \tilde{L}_{J J}(\bfk) \cdot \delta \bfJ(\bfk,\omega)&&
   = T_c\eta_0 k^2 \delta \tilde{\bfJ}_{\perp}(\bfk,\omega).
\end{eqnarray}

Collecting the above results, we arrive at the reduced nonlinear Langevin equation: 
\begin{eqnarray}
-i\omega\tilde{\psi}
   =&&-i k C_{\psi}\delta\tilde{J}_{\parallel}
    -h_c^{-1} i\bfk \cdot \int_{q\Omega} \tilde{\psi}(q)\delta\tilde{\bfJ}(k-q) \nonumber \\
     &&-\lambda_0 k^2 D_{\psi}\chi^{-1}_0(\bfk)\tilde{\psi}+\alpha_1^{-1}\tilde{\theta}_n, \\
-i\omega\tilde{m}
   =&&-\beta_2^{-1} i k \delta \tilde{J}_{\parallel}
    -h_c^{-1}i\bfk\cdot\int_{q\Omega}\tilde{m}(q)\delta\tilde{\bfJ}(k-q), \\
-i\omega \delta\tilde{J}_{\parallel}
   =&&T_c \bl[-ik\chi_0^{-1}(\bfk)C_J \tilde{\psi}-ikC_0^{-1}\beta_2h_c\tilde{m} \nonumber \\
     &&-i k\beta_2^{-1}h_c\gamma_0 \int_{q\Omega}\tilde{\psi}(q)\tilde{\psi}(k-q) \br] \nonumber \\
     &&-k^2\nu_0^l h_c^{-1} \delta\tilde{J}_{\parallel}+\tilde{\theta}_{\parallel}, \label{eq: soundeom} \\
-i\omega\delta\tilde{\bfJ}_{\perp}
   =&&-i T_c {\cal P}_{\perp}(\bfk )\cdot\int_{q\Omega} \bfq 
         \chi_0^{-1}(\bfq) \tilde{\psi}(q)\tilde{\psi}(k-q) \nonumber \\ 
     &&-k^2\eta_0 h_c^{-1}\delta\tilde{\bfJ}_{\perp}+\tilde{\bftheta}_{\perp}, \label{eq: transverse}
\end{eqnarray}
where 
\begin{eqnarray}
D_{\psi} &&\equiv \bl( \frac{n_c T_c}{\alpha_1 h_c}\br)^2, \\
\nu_0^{l} &&\equiv [\zeta_0+2(1-1/d)\eta_0].
\end{eqnarray}
This is the basic equation for the dynamics near the QCD critical point, which is first written down,
and a main result of this paper.

We can compactly rewrite the basic equation in a matrix form:
\begin{equation}
{\cal M}(\bfk,\omega)
  \begin{pmatrix}
  \tilde{\psi}(\bfk ,\omega) \\
  \tilde{m}(\bfk ,\omega) \\
  \delta \tilde{J}_{\parallel}(\bfk,\omega)
  \end{pmatrix}
=\bfV(\bfk,\omega)+\bftheta(\bfk, \omega),
\end{equation}
where
 \begin{equation}
{\cal M}(\bfk, \omega)= 
  \begin{pmatrix}
   -i\omega+\lambda_0 k^2 D_{\psi}\chi^{-1}_0(\bfk) & 0 & i k C_{\psi}  \\
   0 & -i\omega & i k\beta_2^{-1}\\
    i k \chi_0^{-1}(\bfk)C_J T_c &  i k C_0^{-1}\beta_2 h_c T_c & -i\omega +k^2\nu_0^{l} h_c^{-1}
  \end{pmatrix},
\end{equation}
\begin{equation}
\bftheta(\bfk,\omega)= 
  \begin{pmatrix}
  \alpha_1^{-1} \tilde{\theta}(\bfk,\omega)\\
  0 \\
  \tilde{\theta}_{\parallel}(\bfk,\omega)
  \end{pmatrix},
\end{equation}
\begin{equation}
\bfV(\bfk, \omega)= 
  \begin{pmatrix}
  V_{\psi\psi\perp}(\bfk,\omega) +V_{\psi\psi\parallel}(\bfk,\omega) \\
  V_{m m\perp}(\bfk,\omega) +V_{m m\parallel}(\bfk,\omega) \\
  V_{\parallel\psi\psi}(\bfk,\omega)
  \end{pmatrix}, \label{eq: v1}
\end{equation}
and
\begin{eqnarray}
V_{\psi\psi\perp}(\bfk,\omega) &&\equiv
    - h_c^{-1} i\bfk \cdot \int_{q\Omega} \tilde{\psi}(q)\delta\tilde{\bfJ}_{\perp}(k-q), \label{eq: v2 }\\
V_{\psi\psi\parallel}(\bfk,\omega) && \equiv 
    -h_c^{-1} i   \int_{q\Omega} \bfk\cdot(\bfk-\bfq)/|\bfk-\bfq| 
   \nonumber \\ && \times \tilde{\psi}(q)\delta\tilde{J}_{\parallel}(k-q), \\
V_{m m\perp}(\bfk,\omega) &&\equiv -h_c^{-1}i\bfk\cdot\int_{q\Omega}\tilde{m}(q)\delta\tilde{\bfJ}_{\perp}(k-q), \\
V_{m m\parallel}(\bfk,\omega) &&\equiv -h_c^{-1}i k\int_{q\Omega}\tilde{m}(q)\delta\tilde{J}_{\parallel}(k-q), \\
V_{\parallel\psi\psi}(\bfk,\omega) &&\equiv 
\hspace{-0.1cm}-i k T_c\beta_2^{-1}h_c\gamma_0 \int_{q\Omega}\hspace{-0.1cm}
\hspace{-0.2cm}\tilde{\psi}(q)\tilde{\psi}(k-q).\label{eq: v3}
\end{eqnarray}
Because Eq. (\ref{eq: transverse}) is decoupled from the other equations at linear level, 
we do not rewrite it as the matrix form.

Next, we calculate the bare propagator $G^{0}(\bfk ,\omega)={\cal M}^{-1}(\bfk, \omega)$.
The inverse matrix is given as the transposed cofactor matrix divided by $\det {\cal M}$.
The determinant reads
\begin{eqnarray}
\det{\cal M}=&&(-i\omega)^3+(-i\omega)^2 k^2(\lambda D_{\psi}\chi_0^{-1}(\bfk )+\nu_0^l h_c^{-1} ) \nonumber \\
  &&-i\omega k^2(C_0^{-1}h_c T_c+\chi_0^{-1}(\bfk )C_{\psi}C_J T_c) \nonumber \\
  &&+k^4\lambda_0\chi_0^{-1}D_{\psi}C_0^{-1}h_c T_c \nonumber \\
  &&-i\omega k^4\lambda_0\chi_0^{-1}D_{\psi}(\bfk) \nu_0^l h_c^{-1}.
\end{eqnarray}
Here, in the coefficient of  $-i\omega k^2$, 
taking into account the behaviors after renormalization \cite{onukin, glw}, which are
\begin{eqnarray}
C_R^{-1} &&\sim \xi^{-0.2}, \\
\chi_R^{-1}(\bfk ) &&\sim \xi^{-2}+k^2,
\end{eqnarray}
 we neglect $\chi_0^{-1}(\bfk )C_{\psi}C_J T_c$ by comparing with $C_0^{-1}h_c T_c$  .
Then, we can factorize the determinant as 
\begin{eqnarray}
\det {\cal M} &&\sim (-i\omega+\lambda_0(\bfk)\chi_0^{-1}(\bfk)) \nonumber \\
                         &&\times (-i\omega+i k c_s+\frac{1}{2}\nu_0^{l}h_c^{-1} k^2) \nonumber \\
                          && \times             (-i\omega-i k c_s+\frac{1}{2}\nu_0^{l}h_c^{-1} k^2),
\end{eqnarray}
in the long-wavelength region.
Here, we  defined
\begin{eqnarray}
\lambda_0(\bfk)&& \equiv\lambda_0 k^2 D_{\psi}, \\
c_s^2 &&\equiv C_0^{-1}h_c T_c.
\end{eqnarray}
The diagonal components of the cofactor matrix $m$ reads
\begin{eqnarray}
m_{11}\sim &&(-i\omega+i k c_s+\frac{1}{2}\nu_0^l h_c^{-1} k^2)  \nonumber \\
               &&\times (-i\omega-i k c_s+\frac{1}{2}\nu_0^l h_c^{-1} k^2), \\
m_{22}=&&(-i\omega)^2-i\omega k^2\lambda_0\chi_0(\bfk)D_{\psi}\nu_0^l h_c^{-1} \nonumber \\
            &&+k^2\chi_0^{-1}(\bfk)C_{\psi}C_{J}T_c \nonumber \\
            &&+k^4\lambda_0\chi_0(\bfk)D_{\psi}\nu_0^l h_c^{-1} , \\
m_{33}=&&(-i\omega)(-i\omega+\lambda_0(\bfk)\chi_0^{-1}(\bfk)),
\end{eqnarray}
and the off-diagonal components are given by
\begin{eqnarray}
m_{12}&&=k^2\chi_0^{-1}(\bfk)\beta_2^{-1}C_J T_c,\\
m_{13}&&=-k\omega\chi_0^{-1}(\bfk)C_J T_c,\\
m_{21}&&=-k^2 C_0^{-1}h_c C_{\psi}\beta_2 T_c,\\
m_{23}&&=-i k C_0^{-1}h_c\beta_2 T_c(-i\omega+k^2\lambda_0\chi_0(\bfk)D_{\psi}),\\
m_{31}&&=k^2 C_0^{-1}h_c C_\psi\beta_2 T_c,\\
m_{32}&&=-i k\beta_2^{-1}(-i\omega+\lambda_0 (\bfk )\chi_0^{-1}(\bfk )).
\end{eqnarray}
Here, we neglect the off-diagonal components 
because they would not yield dominant contributions to the transport coefficients.
Then, we obtain the bare propagator as
\begin{equation}
G^0(\bfk, \omega)= 
  \begin{pmatrix}
  G^0_{\psi}(\bfk,\omega) & 0 & 0 \\
  0 & G^0_{m}(\bfk,\omega) & 0\\
  0 & 0 & G^0_{\parallel}(\bfk,\omega)
  \end{pmatrix}.\label{eq: g0}
\end{equation}
with
\begin{eqnarray}
G^0_{\psi}(\bfk,\omega)  =&&\frac{1}{-i\omega+\lambda(\bfk)\chi_0^{-1}(\bfk)}, 
\label{eq: gthermal}\\
G^0_{\parallel}(\bfk,\omega) 
  \sim &&\frac{1}{2}\bl[ \frac{1}{-i\omega+i k c_s+\frac{1}{2}\nu_0^l h_c^{-1}k^2} \nonumber \\
                         &&+\frac{1}{-i\omega-i k c_s+\frac{1}{2}\nu_0^{l}h_c^{-1} k^2}\br].
\label{eq: gsound}
\end{eqnarray}
$G^0_{mm}(\bfk,\omega)$ is not needed in later calculations.
The bare propagator of $\delta \bfJ_{\perp}$ is trivially given by 
\begin{equation}
G^0_{\perp}(\bfk,\omega)=\frac{1}{-i\omega+\eta_0 k^2 h_c^{-1}}.
\label{eq: gviscous}
\end{equation}
We finally arrive at the equations of motion as the self-consistent form:
\begin{equation}
  \begin{pmatrix}
  \tilde{\psi}(\bfk ,\omega) \\
  \tilde{m}(\bfk ,\omega) \\
  \delta \tilde{J}_{\parallel}(\bfk,\omega)
  \end{pmatrix}
=
 \begin{pmatrix}
  \tilde{\psi}^0(\bfk ,\omega) \\
  0 \\
  \delta \tilde{J}_{\parallel}^0(\bfk,\omega)
  \end{pmatrix}
+G^0(\bfk,\omega)\bfV(\bfk,\omega),
\label{eq: selfeom}
\end{equation}
and
\begin{equation}
\delta \tilde{\bfJ}_{\perp}(\bfk,\omega)
=\delta \tilde{\bfJ}_{\perp}^0(\bfk,\omega)+G^0_{\perp}\bfV_{\perp \psi\psi}(\bfk,\omega),
\label{eq: selfjeom}
\end{equation}
where
\begin{eqnarray}
\tilde{\psi}^0(\bfk ,\omega) &&=G^0_{\psi}(\bfk,\omega)  \alpha_1^{-1} \tilde{\theta}_n(\bfk,\omega),
\label{eq: a}\\
\delta \tilde{J}_{\parallel}^0(\bfk,\omega) &&=G^0_{\parallel}(\bfk,\omega) \tilde{\theta}_{\parallel}(\bfk,\omega),\\
\delta \tilde{\bfJ}_{\perp}^0(\bfk,\omega) &&=G^0_{\perp}(\bfk,\omega)\tilde{\bftheta}_{\perp}(\bfk,\omega), \\
\bfV_{\perp \psi\psi}(\bfk,\omega)
=&&-i T_c{\cal P}_{\perp}(\bfk)\cdot\int_{q\Omega}\bfq \chi_0^{-1}(\bfq) 
   \tilde{\psi}(q)\tilde{\psi}(k-q). \label{eq: v4} 
\end{eqnarray}
Here, $\tilde{\psi}^0(\bfk ,\omega) $, $\delta \tilde{J}_{\parallel}^0(\bfk,\omega) $
and $\delta \tilde{\bfJ}_{\perp}^0 $ are the bare variables that are the solutions without the nonlinear terms.
Iterating Eqs. (\ref{eq: selfeom}) and (\ref{eq: selfjeom}),
we can obtain perturbative expansions about nonlinear interactions $\bfV$ and $\bfV_{\perp \psi\psi}$.
We note that the first and third rows of Eq. (\ref{eq: selfeom}) 
are the equations of motion for the thermal and sound modes, respectively, 
while Eq. (\ref{eq: selfjeom}) is for the viscous mode.
We also stress that Eqs. (\ref{eq: gthermal})-(\ref{eq: gviscous}) are the propagators of
the thermal, sound and viscous modes, respectively.    

Now, we calculate the two body correlation of 
$\tilde{\psi}^0(\bfk ,\omega)$ and $\delta \tilde{\bfJ}_{\perp}^0(\bfk,\omega) $,
which are needed in later calculations.
\begin{eqnarray}
\la\tilde{\psi}^0(\bfk_1 ,\omega_1)\tilde{\psi}^0(\bfk_2 ,\omega_2) \ra
  =&&G^0_{\psi}(\bfk_1 ,\omega_1)G^0_{\psi}(\bfk_2 ,\omega_2)\alpha_1^{-2} \nonumber \\
    &&\times \la\tilde{\theta}(\bfk_1,\omega_1)\tilde{\theta}(\bfk_2,\omega_2) \ra.
\end{eqnarray}
Using the fluctuation dissipation relation Eq.(\ref{eq: fdrn}), we find
\begin{equation}
\la\tilde{\theta}(\bfk_1,\omega_1)\tilde{\theta}(\bfk_2,\omega_2) \ra
=2  \alpha_1^2 \lambda_0(\bfk)(2\pi)^{d+1}\delta( k_1+k_2),
\end{equation}
and
\begin{eqnarray}
\la\tilde{\psi}^0(\bfk_1 ,\omega_1)\tilde{\psi}^0(\bfk_2 ,\omega_2) \ra
=&&\frac{2  \lambda_0(\bfk_1)}{\omega_1^2+\lambda_0^2(\bfk_1)\chi_0^{-2}(\bfk_1)} \nonumber \\
&&\hspace{-0.1cm}\times (2\pi)^{d+1}\delta( k_1+k_2),
\label{eq: pcorrelation}
\end{eqnarray}
where $\delta( k_1+k_2) \equiv \delta( \bfk_1+\bfk_2)\delta( \omega_1+\omega_2)$.
By a similar calculation, we have
\begin{eqnarray}
\la \delta\tilde{J}_{\perp}^{i}(\bfk_1,\omega_1) \delta\tilde{J}_{\perp}^{i}(\bfk_2,\omega_2) \ra
&&=\frac{2 T_c \eta_0(\bfk_1)}{\omega_1^2+\eta_0^2(\bfk_1)h_c^{-2}}({\cal P}_{\perp}(\bfk_1))_{i j} \nonumber \\
&&\times (2\pi)^{d+1}\delta( k_1+k_2),
\end{eqnarray}
where $\eta_0(\bfk)=\eta_0 k^2$.
For a later convenience,
we define 
\begin{eqnarray}
C^0_{\psi}(\bfk, \omega ) = &&\frac{2  \lambda_0(\bfk )}{\omega^2+\lambda_0^2(\bfk )\chi_0^{-2}(\bfk )}, \\
C^0_{\perp}(\bfk, \omega )= &&\frac{2 T_c \eta_0(\bfk )}{\omega^2+\eta_0^2(\bfk )h_c^{-2}}.
\end{eqnarray}

\chapter{Renormalization of the thermal and viscous modes}

\label{sec: RG1}
\begin{figure*}[!t]
    \begin{minipage}{1.0\hsize}
     \centering
     \includegraphics[width=\hsize]{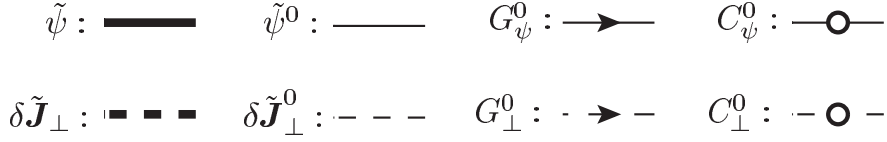}
       \caption{Diagrams for the full and bare variables, 
                   the bare propagators and the bare correlations.}
     \label{fig: difinition}
    \end{minipage}
\end{figure*}

\begin{figure}[!t]
    \begin{minipage}{1.0\hsize}
     \centering
     \includegraphics[width=\hsize]{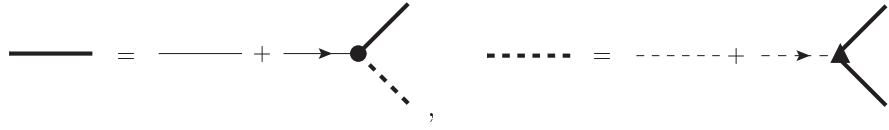}
       \caption{Diagrams of the equations of motion for the thermal and viscous modes.
                   The left and right hand side respectively denote Eqs. (\ref{eq: selfpeom}) and (\ref{eq: selfjeom}). }
     \label{fig: eom}
    \end{minipage}
\end{figure}

Here, we first derive the RG equations for the thermal conductivity and the shear viscosity.
Now, we note that the sound mode is not a genuine-relevant mode but a secondary mode 
that is strongly affected by order-parameter fluctuations 
but yields only a negligible feedback for the order parameters \cite{Minami:2009hn, kroll}.
Then, we can neglect the sound mode for the minimal critical dynamics;
however, the bulk viscosity is not renormalized in that case.
Here, to first analyze the minimal  dynamics, 
we neglect the secondary mode, which is renormalized in the next section. 
In that case, the equations of motion are 
\begin{eqnarray}
\tilde{\psi}(\bfk, \omega) =&&\tilde{\psi}^{0}(\bfk, \omega)+G_{\psi}^{0}(\bfk, \omega )V_{\psi\psi\perp}(\bfk, \omega )
\label{eq: selfpeom}
\end{eqnarray}
and Eq. (\ref{eq: selfjeom}).
For a diagrammatic treatment ,
we denote the full and bare variables, the bare propagators 
and the bare correlation functions as Fig. \ref{fig: difinition}. 
Then, we can represent the equations of motion (\ref{eq: selfpeom}) and (\ref{eq: selfjeom}) as Fig. \ref{fig: eom}.

\begin{figure*}[!t]
    \begin{minipage}{1.0\hsize}
     \centering
     \includegraphics[width=\hsize]{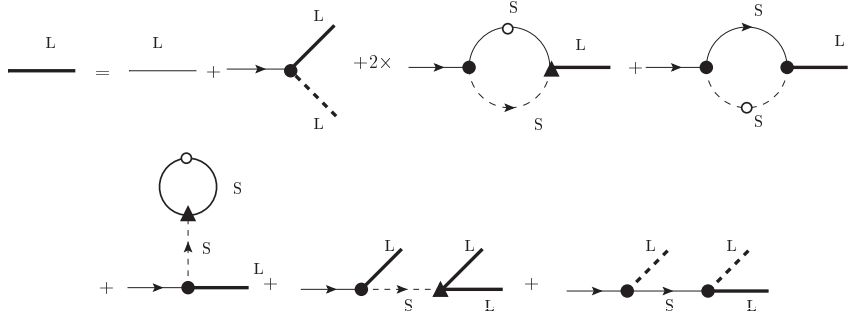}
       \caption{Diagrams for the coarse-grained equation of motion for $\psi$.
                   The letters, L and S, respectively denote the long- and short-wavelength components
                   (see the text below Eq.(\ref{eq: abcde})).}
     \label{fig: cgpeom}
    \end{minipage}
\end{figure*}

For  coarse gaining, we decompose the variables into the long- 
and short-wavelength components as
\begin{equation}
\tilde{\psi}(\bfk, \omega) = \tilde{\psi}^{\rm L} (\bfk, \omega )+\tilde{\psi}^{\rm S}(\bfk, \omega), 
\end{equation}
with
\begin{eqnarray}
\tilde{\psi}^{\rm L}(\bfk, \omega ) \equiv&& \Theta(\Lambda-\delta\Lambda-k)\tilde{\psi}(\bfk, \omega),\\  
\tilde{\psi}^{\rm S} (\bfk, \omega)\equiv&& \Theta(k-\Lambda-\delta\Lambda)\tilde{\psi}(\bfk, \omega) ,
\label{eq: abcde}
\end{eqnarray}
where $\Theta(x)$ is a step function; 
i.e., the wavenumber is decomposed into $0 < k < \Lambda - \delta \Lambda$ 
and $\Lambda - \delta \Lambda < k < \Lambda$. 
Hereafter, quantities with the suffixes L and S are supposed to be decomposed as above.
To average over the $\tilde{\psi}^{0 {\rm S}}$ and $\delta \tilde{\bfJ}_{\perp}^{0 {\rm S}}$,
we must solve the equation of motion about them.  
Here, we solve the equations to second order in the nonlinear interactions
and average over $\tilde{\psi}^{0 {\rm S}} $ and $\delta \tilde{\bfJ}_{\perp}^{0 {\rm S}}$.
Then, we find the coarse-grained equation of motion for $ \psi$, 
which is diagrammatically given by Fig. \ref{fig: cgpeom}.
The last two terms in Fig. \ref{fig: cgpeom} represent nonlinear interactions being of third order, 
and can be neglected.
Furthermore, the fifth term vanishes due to the relation between the step and delta functions in the loop integral.
\begin{figure*}[!t]
    \begin{minipage}{1.0\hsize}
     \centering
     \includegraphics[width=\hsize]{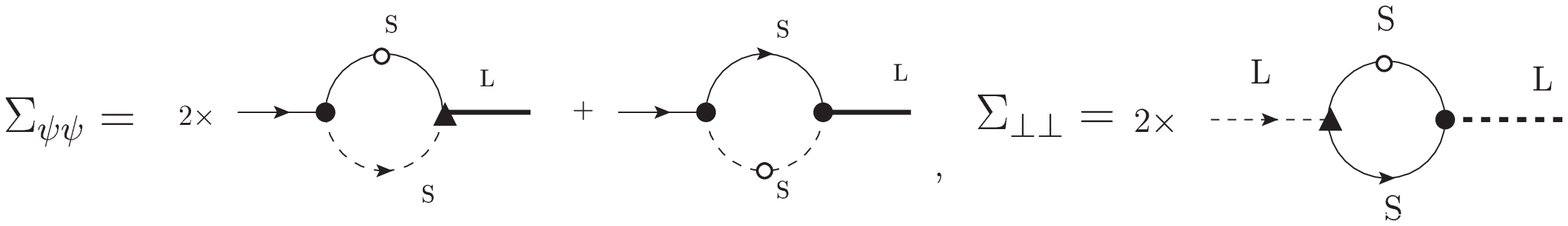}
       \caption{Diagrams for the self energies.}
     \label{fig: self}
    \end{minipage}
\end{figure*}%
Introducing the self energy $\Sigma_{\psi\psi}$, which is graphically represented in Fig. \ref{fig: self},
we can write the coarse-grained equation for $ \psi$ as
\begin{eqnarray}
\tilde{\psi}^{\rm L}(\bfk, \omega) 
  =&&\tilde{\psi}^{0{\rm L}}(\bfk, \omega)+G_{\psi}^{0{\rm L}}(\bfk, \omega )V_{\psi\psi\perp}^{\rm L} (\bfk, \omega )
\nonumber \\
   &&+\tilde{\psi}^{\rm L}(\bfk, \omega)G_{\psi}^{0{\rm L}}(\bfk, \omega )\Sigma_{\psi\psi}(\bfk, \omega ).
   \label{eq: cgpeom}
\end{eqnarray} 
The self energy is  
\begin{eqnarray}
\Sigma_{\psi\psi}&&(\bfk, \omega ) =-T_c h_c^{-1} k^2 \chi_0^{-1} (\bfk)  \nonumber \\
  &&\hspace{-0.3cm}\times\int_q  \frac{(\hat{\bfk}\cdot{\cal P}(\bfk-\bfq )\cdot\hat{\bfk})\chi_0(\bfq)}
  {-i\omega+\lambda_0(\bfq)\chi_0^{-1}(\bfq)+\eta_0(\bfk-\bfq)h_c^{-1} }, \label{eq:aaaaa}
\end{eqnarray}
where $\eta_0(\bfk)=\eta_0 k^2$.
Solving Eq. (\ref{eq: cgpeom}) about $\tilde{\psi}^{\rm L}$,  we have
\begin{eqnarray}
\tilde{\psi}^{\rm L}=&&[(G_{\psi}^{0{\rm L}})^{-1} -\Sigma_{\psi\psi}]^{-1}\alpha_1^{-1}\tilde{\theta}_n
 \nonumber \\ &&
+[(G_{\psi}^{0{\rm L}})^{-1} -\Sigma_{\psi\psi}]^{-1}V_{\psi\psi\perp}^{\rm L}.
\end{eqnarray}
where we used Eq. (\ref{eq: a}).
Introducing renormalized variables as
\begin{eqnarray}
(G_{\psi {\rm R}})^{-1}(\bfk, \omega )
  =&&(G_{\psi}^{0 {\rm L} })^{-1}(\bfk, \omega ) -\Sigma_{\psi \psi}(\bfk, \omega ), \label{eq: rppro}\\
\tilde{\psi}^{0{\rm L}}_ {\rm R}(\bfk, \omega) =&&G_{\psi{\rm R}}(\bfk, \omega )\alpha_1^{-1}\tilde{\theta}_n(\bfk, \omega ), 
\end{eqnarray}
we can rewrite Eq.(\ref{eq:aaaaa}) as the renormalized equation of motion:
\begin{eqnarray}
\tilde{\psi}^{\rm L}=&&\tilde{\psi}^{0{\rm L}}_ {\rm R}(\bfk, \omega )
+G_{\psi {\rm R}}(\bfk, \omega )V_{\psi\psi\perp}^{\rm L}.
\label{eq: bbb}
\end{eqnarray}
We now require that the renormalized propagator has the same form as the bare one:
\begin{equation}
(G_{\psi {\rm R}})^{-1}(\bfk, \omega )= -i\omega +\lambda_R D_\psi k^2 \chi_0^{-1}(\bfk ),
\end{equation} 
where $\lambda_{\rm R}$ is the renormalized thermal conductivity.
That is, we require that the only transport coefficients are explicitly renormalized.
The small correction for the thermal conductivity $\delta \lambda \equiv \lambda_{\rm R}-\lambda_0$ reads
\begin{eqnarray}
\delta \lambda =&&-\lim_{k, \omega \to 0}[(D_\psi k^2 \chi_0(\bfk ))^{-1}\Sigma_{\psi \psi}(\bfk, \omega )],
                                                                                                                            \nonumber \\
                    =&&\frac{T_c}{ h_c D_\psi }\int_q 
                         \frac{(\hat{\bfk}\cdot{\cal P}(\bfq )\cdot\hat{\bfk})\chi_0(\bfq)}
                           {\lambda_0(\bfq)\chi_0^{-1}(\bfq)+\eta_0(\bfq)h_c^{-1} }.        
\end{eqnarray}
We  approximate the denominator and the numerator as
\begin{eqnarray}
\lambda_0(\bfq)\chi_0^{-1}(\bfq)&&+\eta_0(\bfq)h_c^{-1} \sim  \eta_0 (\bfk) h_c^{-1}, \\
\chi_0^{-1}(\bfq ) = && r_0 + q^2 \sim q^2, 
\end{eqnarray}
near the critical point \cite{siggia}.
Then, we find
\begin{eqnarray}
\delta \lambda &&\sim  \frac{T_c}{D_\psi \eta_0} 
     \int \frac{d\Omega_d}{(2\pi)^d}(\hat{\bfk}\cdot{\cal P}(\bfq )\cdot\hat{\bfk})
     \int^{\Lambda}_{\Lambda-\delta \Lambda} d q q^{d-5}   \nonumber \\
  &&=- \frac{T_c}{D_\psi \eta_0} \int \frac{d\Omega_d}{(2\pi)^d}(\hat{\bfk}\cdot{\cal P}(\bfq )\cdot\hat{\bfk}) 
     \Lambda^{d-5} \delta \Lambda,
\end{eqnarray}
where $d\Omega_d$ is the solid angle in the space dimension $d$.
Therefore, we obtain the RG equation for the thermal conductivity:
\begin{equation}
-\Lambda \frac{\partial \lambda}{\partial \Lambda}=\frac{T_c}{D_\psi \eta(\Lambda)} \int \frac{d\Omega_d}{(2\pi)^d}(\hat{\bfk}\cdot{\cal P}(\bfq )\cdot\hat{\bfk}) 
     \Lambda^{d-4}, 
\end{equation}
where $\eta_0$ is rewritten as $\eta (\Lambda)$. 
For the space dimensions, $d=4-\epsilon$,
the angle integral is given by 
\begin{equation}
 \int \frac{d\Omega_4}{(2\pi)^4}(\hat{\bfk}\cdot{\cal P}(\bfq )\cdot\hat{\bfk})  =\frac{3}{4}K_4,
\end{equation}
where $K_4$ is the surface area of a unit sphere in 4 dimensions divided by $(2\pi)^4$.
The RG equation in $4- \epsilon$ dimensions reads
\begin{equation}
-\Lambda \frac{\delta \lambda}{\delta \Lambda}=\frac{3}{4} f(\Lambda) \lambda(\Lambda), 
\label{eq: rgl2}
\end{equation}
where we introduced
\begin{equation}
f(\Lambda) \equiv \frac{T_c K_4}{D_\psi \eta(\Lambda) \lambda(\Lambda) \Lambda^{\epsilon}},
\label{eq: flambda}
\end{equation} 
for a later convenience.

By making coarse graining of the viscous mode with a similar procedures as above,
we obtain a small correction for the shear viscosity:
\begin{equation} 
\delta \eta = -\lim_{k, \omega \to 0}[(k^2 h_c^{-1}(d-1))^{-1}\sum_i(\Sigma_{\perp\perp}(\bfk, \omega))_{ii}],
\label{eq: deltaeta}
\end{equation} 
where $(\Sigma_{\perp\perp}(\bfk, \omega))_{ij}$ is the self energy for the viscous mode and given by
\begin{eqnarray}
(\Sigma_{\perp\perp}&&(\bfk, \omega))_{i j}=-T_c h_c^{-1}
\int_q \chi_0(\bfk-\bfq) ({\cal P}_{\perp}(\bfk )\cdot \bfq )_i q_j \nonumber \\
&& \hspace{-1.0cm} \times\frac{\chi_0^{-1}(\bfq) -\chi_0^{-1}(\bfk-\bfq)}
{-i\omega +\lambda_0(\bfq )\chi_0^{-1}(\bfq )+\lambda(\bfk-\bfq)\chi_0^{-1}(\bfk-\bfq )},
\label{eq: jself}
\end{eqnarray}
which is diagrammatically represented as Fig. \ref{fig: self} .
In the space dimension $d=4-\epsilon$,
we find the RG equation for the shear viscosity
\begin{equation}
-\Lambda \frac{\partial \eta (\Lambda)}{\partial \Lambda}=\frac{1}{24} f(\Lambda) \eta(\Lambda),
\label{eq: rge2}
\end{equation}
where the prefactor $1/24$ comes from the angular integral in Eq. (\ref{eq: jself})
 and the factor $(d-1)^{-1}$ in Eq. (\ref{eq: deltaeta}).
 
Differentiating Eq. (\ref{eq: flambda}) about $\Lambda$,
we have the RG equation for $f(\Lambda)$
\begin{equation}
-\Lambda \frac{\partial f(\Lambda)}{\partial \Lambda}=(\epsilon-\frac{19}{24}f(\Lambda ))f(\Lambda ).
\end{equation} 
 
\chapter{Renormalization of the sound mode}

\label{sec: RG2}
Next, let us make a coarse graining of the sound mode for the renormalized bulk viscosity.   
Because a feedback from the sound mode is neglected, 
we must renormalize the mode with a method separating 
relevant and secondary modes \cite{kroll}.
Here, we take the method developed by Onuki \cite{onukin, onuki} 
, in which RG equations are derived from fluctuation-dissipation relations.

Now, we consider the equation of motion for the sound mode, (\ref{eq: soundeom}):
\begin{eqnarray}
-i\omega\delta\tilde{J}_{\parallel}
   =&&- i k T_c \bl[ \chi_0^{-1}(\bfk)C_J \tilde{\psi}+ C_0^{-1}\beta_2 h_c\tilde{m} \nonumber \\
     &&+\beta_2^{-1}h_c\gamma_0 \int_{q\Omega}\tilde{\psi}(q)\tilde{\psi}(k-q) \br] \nonumber \\
     &&-k^2\nu_0^l h_c^{-1}\delta\tilde{J}_{\parallel} +\tilde{\theta}^{0}_{\parallel}, 
     \label{eq: soundeom2}
\end{eqnarray}
where the noise term $\tilde{\theta}_{\parallel}^0$ satisfies the fluctuation dissipation relation:
\begin{eqnarray}
\la \tilde{\theta}^{0}_{\parallel}(\bfk_1, \omega_1) \tilde{\theta}^{0}_{\parallel}(\bfk_2, \omega_2) \ra &&
=2 T_c k_1^2 \nu_0^{l} \nonumber \\ &&
\times (2\pi)^{d+1}\delta(k_1+k_2).
\end{eqnarray}
Since $\delta \bfJ_{\parallel}$ is a conserved density projected onto $\hat{\bfk}$,
we can rewrite  Eq. (\ref{eq: soundeom2}) as 
\begin{equation}
-i\omega \delta\tilde{J}_{\parallel} (\bfk, \omega) =i \bfk \cdot \tilde{\Pi}(\bfk, \omega )\cdot \hat{\bfk},
\end{equation}
where $\tilde{\Pi}_{i j}$ is the stress tensor.
If we take  $z$ direction as  $\hat{\bfk}$,
$\tilde{\Pi}_{z z}$ reads
\begin{eqnarray}
\tilde{\Pi}_{z z}(\bfk, \omega ) &&= -T_c \bl[\chi_0^{-1}(\bfk)C_J \tilde{\psi}(\bfk, \omega)
                                            + C_0^{-1}\beta_2 h_c\tilde{m} (\bfk, \omega)\nonumber \\
     &&+\beta_2^{-1}h_c\gamma_0 \int_{q\Omega}\tilde{\psi}(q)\tilde{\psi}(k-q) \br] \nonumber \\
     &&+i k\nu_0^l h_c^{-1}\delta\tilde{J}_{\parallel}(\bfk, \omega) +\tilde{\pi}^{0}_{z z}(\bfk, \omega) ,
\end{eqnarray} 
where $\tilde{\pi}^{0}_{i j}(\bfk, \omega)$ is the random-stress tensor coming from microscopic process
and satisfies the relation, 
$i \bfk \cdot \tilde{\pi}^{0}(\bfk, \omega) \cdot \hat{\bfk}=\tilde{\theta}^{0}_{\parallel}(\bfk, \omega)$.

We now consider how Eq. (\ref{eq: soundeom2}) is affected by the coarse-graining procedure.
In the coarse-graining procedure,
 the variables, $\tilde{\psi}^{\rm S}$, $\tilde{m}^{\rm S}$ and $\delta \tilde{\bfJ}^{\rm S}$ 
are eliminated from Eq. (\ref{eq: soundeom2}).
The eliminated variables do not disappear from the equation of motion but are
implicitly contained in the noise term.
In other words, we convert the macroscopic process in the wavenumber shell
$\Lambda - \delta \Lambda < k < \Lambda$ into the microscopic process.
In this procedure, the noise term is implicitly renormalized as follows
\begin{equation}
\tilde{\theta}^{R}_{\parallel}(\bfk, \omega)=\tilde{\theta}^{0}_{\parallel}(\bfk, \omega)
+\tilde{\theta}^{\rm Macro}_{\parallel}(\bfk, \omega) \label{eq: macro},
\end{equation}
where
\begin{eqnarray}
\tilde{\theta}^{\rm Macro}_{\parallel}(\bfk, \omega )
   &&\equiv  i\bfk \cdot \tilde{\pi}^{\rm Macro}(\bfk, \omega ) \cdot \hat{\bfk}, \\
\tilde{\pi}^{\rm Macro}_{z z}(\bfk, \omega )&&\equiv -T_c \bl[\chi_0^{-1}(\bfk)C_J \tilde{\psi}^{\rm S}
                                            + C_0^{-1}\beta_2 h_c\tilde{m}^{\rm S} \nonumber \\
     &&+\beta_2^{-1}h_c\gamma_0 \int_{q\Omega}\tilde{\psi}^{\rm S}(q)\tilde{\psi}^{\rm S}(k-q) \br] 
      +i k\nu_0^l h_c^{-1}\delta\tilde{J}_{\parallel}^{\rm S}, \label{eq: pi}\\
     \sim && -T_c \beta_2^{-1}h_c\gamma_0 \int_{q\Omega}\tilde{\psi}^{\rm S}(q)\tilde{\psi}^{\rm S}(k-q),
\end{eqnarray}
where we neglect the linear terms in Eq. (\ref{eq: pi}) that is irrelevant for the following argument.
The new term $\tilde{\theta}^{\rm Macro}_{\parallel}(\bfk, \omega )$, being due to the coarse graining,
contributes the transport coefficient through the fluctuation-dissipation relation:
\begin{eqnarray}
\la &&\tilde{\theta}^{\rm Macro}_{\parallel}(\bfk_1, \omega_1) 
   \tilde{\theta}^{\rm Macro}_{\parallel}(\bfk_2, \omega_2) \ra \nonumber \\ 
         &&=2 T_c k_1^2 \delta\nu^{l}(\bfk_1, \omega_1 )(2\pi)^{d+1}\delta(k_1+k_2),
         \label{eq: bb}
\end{eqnarray}
where we have assumed that the renormalized equation of motion has the same form as Eq. (\ref{eq: soundeom2}).
We note that this assumption is equivalent to the requirement below Eq. (\ref{eq: bbb}).
Now, we calculate the left-hand side in Eq. (\ref{eq: bb}):
\begin{eqnarray}
\la \tilde{\theta}^{\rm Macro}_{\parallel}(\bfk_1, \omega_1) 
\tilde{\theta}^{\rm Macro}_{\parallel}(\bfk_2, \omega_2) \ra
=&&-k_1 k_2 (T_c h_c \beta_2^{-1})^2\gamma_0^{2} \nonumber \\
&&\hspace{-4.5cm}
\times  \int_{q_1 \Omega_1 q_2 \Omega_2} \hspace{-1.0cm}
\la \tilde{\psi}^{\rm S}(q_1) \tilde{\psi}^{\rm S}(k_1-q_1)
\tilde{\psi}^{\rm S}(q_2) \tilde{\psi}^{\rm S}(k_2-q_2) \ra.
\end{eqnarray}
Approximating the variable by the bare one, $\tilde{\psi}^{\rm S} \sim \tilde{\psi}^{0 {\rm S}}$,
we find
\begin{eqnarray}
\la && \tilde{\theta}^{\rm Macro}_{\parallel}(\bfk_1, \omega_1) 
     \tilde{\theta}^{\rm Macro}_{\parallel}(\bfk_2, \omega_2) \ra
     = (2\pi)^{d+1}\delta( k_1+k_2) \nonumber \\
  && \times 2 k_1^2 (T_c h_c \beta_2^{-1})^2\gamma_0^{2} 
     \int_{q \Omega}C_{\psi}^{0{\rm S}}(q) C_{\psi}^{0{\rm S}}(k_1-q),
\end{eqnarray}
where we have used Eq. (\ref{eq: pcorrelation}) and neglected a term corresponding to a disconnected diagram.
Then, comparing with Eq. (\ref{eq: bb}), we obtain the correction to the longitudinal-kinetic viscosity:
\begin{equation}
\delta \nu^{l}(\bfk, \omega ) =T_c \beta_2^{-2} h_c^2 \gamma_0^2
\int_{q \Omega}  C_{\psi}^{0{\rm S}}(q) C_{\psi}^{0{\rm S}}(k-q).
\end{equation}
We are not interested in the frequency- or wavenumber-dependent bulk viscosity
and then take the limit $k, \omega \rightarrow 0$:
\begin{eqnarray}
\delta \nu^{l} &&\equiv \lim_{k, \omega \to 0}\delta \nu^{l}(\bfk, \omega ) \nonumber \\
 &&=T_c \beta_2^{-2} h_c^2 \gamma_0^2
\int_{q \Omega}  \bl( C_{\psi}^{0{\rm S}}(q) \br)^2 .
\end{eqnarray}
After the integration, we find the RG equation for longitudinal kinetic viscosity:
\begin{eqnarray}
-\Lambda\frac{\partial \nu^l(\Lambda)}{\partial \Lambda}
=\frac{T_c h_c^2 K_4}{\beta_2^2 D_\psi}\gamma^2(\Lambda)\lambda^{-1}(\Lambda)\Lambda^{-\epsilon-4}.
\end{eqnarray} 
where we have rewritten the static parameter $\gamma_0$ as $\gamma (\Lambda)$ to 
denote its cutoff dependence as mentioned in the text.
The asymptotic behavior obtained from this RG equation is different from 
the shear viscosity's behavior, so we  replace above RG equation as 
 \begin{eqnarray}
-\Lambda\frac{\partial \zeta(\Lambda)}{\partial \Lambda}
=\frac{T_c h_c^2 K_4}{\beta_2^2 D_\psi}\gamma^2(\Lambda)\lambda^{-1}(\Lambda)\Lambda^{-\epsilon-4}.
\end{eqnarray} 

Although, by this method,
we could more easily obtain the RG equations for the thermal conductivity and shear viscosity,
we have taken the diagrammatic method for an instructive purpose.

\end{document}